\documentclass[a4paper,11pt]{article}

\usepackage{jcappub}
\usepackage[T1]{fontenc} 
\usepackage{color}

\def\l{\left}
\def\r{\right}
\def\ddd{\mathrm{d}}
\def\be#1\ee{\begin{equation}#1\end{equation}}
\def\bl#1\el{\begin{align}#1\end{align}}
\def\ba#1\ea{\begin{align*}#1\end{align*}}
\def\nn{\nonumber}

\title{Growth of curvature perturbations for PBH formation \& detectable GWs in non-minimal curvaton scenario revisited}

\author[a]{Chao Chen}
\author[b]{Anish Ghoshal}
\author[b]{Zygmunt Lalak}
\author[c,d]{Yudong Luo}
\author[e]{Abhishek Naskar}
 
\affiliation[a]{Jockey Club Institute for Advanced Study, The Hong Kong University of Science and Technology, Hong Kong S.A.R., China}
\affiliation[b]{Institute of Theoretical Physics, Faculty of Physics, University of Warsaw, \\ ul. Pasteura 5, 02-093 Warsaw, Poland}
\affiliation[c]{School of Physics, Peking University, Beijing 100871, China}
\affiliation[d]{Kavli Institute for Astronomy and Astrophysics, Peking University, Beijing 100871, China}
\affiliation[e]{Department of Physics, Indian Institute of Technology Bombay, Mumbai, 400076, India}

\emailAdd{iascchao@ust.hk}
\emailAdd{anish.ghoshal@fuw.edu.pl}
\emailAdd{zygmunt.lalak@fuw.edu.pl}
\emailAdd{yudong.luo@pku.edu.cn}
\emailAdd{abhiatrkmrc@gmail.com}

\abstract{
	We revisit the growth of curvature perturbations in non-minimal curvaton scenario with a non-trivial field metric $\lambda(\phi)$ where $\phi$ is an inflaton field, and incorporate the effect from the non-uniform onset of curvaton's oscillation in terms of an axion-like potential. 
	The field metric $\lambda(\phi)$ plays a central role in the enhancement of curvaton field perturbation $\delta\chi$, serving as an effective friction term which can be either positive or negative, depending on the first derivative $\lambda_{,\phi}$.
	Our analysis reveals that $\delta\chi$ undergoes the superhorizon growth when the condition $\eta_\text{eff} \equiv - 2 \sqrt{2\epsilon} M_\text{Pl} { \lambda_{,\phi} \over \lambda} < -3$ is satisfied. 
	This is analogous to the mechanism responsible for the amplification of curvature perturbations in the context of ultra-slow-roll inflation, namely the growing modes dominate curvature perturbations. 
	As a case study, we examine the impact of a Gaussian dip in $\lambda(\phi)$ and conduct a thorough investigation of both the analytical and numerical aspects of the inflationary dynamics.
	Our findings indicate that the enhancement of curvaton perturbations during inflation is not solely determined by the depth of the dip in $\lambda(\phi)$. Rather, the first derivative $\lambda_{,\phi}$ also plays a significant role, a feature that has not been previously highlighted in the literature.
	Utilizing the $\delta \mathcal{N}$ formalism, we derive analytical expressions for both the final curvature power spectrum and the non-linear parameter $f_\text{NL}$ in terms of an axion-like curvaton's potential leading to the non-uniform curvaton's oscillation.
	Additionally, the resulting primordial black hole abundance and scalar-induced gravitational waves are calculated, which provide observational windows for PBHs.
}


\arxivnumber{2305.12325}
	
\begin{document}
	
\maketitle
\flushbottom

\section{Introduction}

The cosmic microwave background radiation (CMB) experiments have already confirmed that the primordial curvature perturbation $\zeta$ is nearly scale-invariant and Gaussian on large scales (around the pivot scale $k_\text{pivot} = 0.05$ Mpc$^{-1}$), with the amplitude of its power spectrum $\mathcal{P}_\zeta \sim 10^{-9}$ \cite{Planck:2018jri}. However, the small-scale primordial curvature perturbations are free from the current observational constraints (see Fig.~\ref{fig:totpzeta}), and tremendous efforts have been dedicated to the theoretical and phenomenological studies of the potential growth of the small-scale primordial curvature perturbations in past decades. One of the most promising phenomena is the formation of primordial black holes (PBHs) through the direct gravitational collapse of overdense regions in the early Universe \cite{Hawking:1971ei, Carr:1974nx, Carr:1975qj}.
The fact that PBHs can exist with different masses and result in various astrophysical and cosmological phenomena, has captured considerable research interest.
For example, detecting Hawking radiation may be possible using light PBHs with a mass of less than $10^{-17} M_\odot$\footnote{$M_\odot \simeq 2 \times 10^{33} \text{~g}$ is the solar mass.} \cite{Carr:2009jm, Carr:2020gox, Luo:2020dlg, Cai:2021zxo, Cai:2021fgm}. Heavy PBHs with a mass of around $10^{5} M_\odot$ could serve as seed black holes for supermassive black holes \cite{Bean:2002kx, Kawasaki:2012kn, Yuan:2023bvh, Cai:2023ptf}. Moreover, the medium-sized PBHs with a mass in the tens of solar masses could be candidates for LIGO/Virgo GW events \cite{Bird:2016dcv, Sasaki:2016jop, Jedamzik:2020ypm}. It is worth noting that the possibility of PBHs serving as a candidate for dark matter has been a topic of discussion for several decades. However, tighter experimental constraints increasingly challenge their ability to account for dark matter \cite{Carr:2020xqk, Green:2020jor, Escriva:2022duf}.

Extensive theoretical research has been dedicated to studying PBH formation in vast models of the early Universe, especially in the context of the inflationary models (refer to comprehensive reviews \cite{Sasaki:2016jop, Escriva:2022duf} for the summary of PBH formation models). In this paper, we revisit the growth of curvature perturbations in a non-minimal curvaton scenario proposed recently by Ref. \cite{Pi:2021dft} and studied further by the following Ref. \cite{Meng:2022low} with three specific forms of field metric, including the Gaussian-like, rectangular and oscillating dips. However, some discrepancies were observed in the curvaton power spectrum at the end of inflation, for which the underlying reason is unclear. 
Our paper aims to investigate the details of curvaton field perturbations during inflation, regarding a non-trivial field metric identified as an effective friction term [cf. Eq. \eqref{inf:etaeff}]. The growth of curvaton field perturbations occurs when this effective friction term becomes negative enough [cf. Eq. \eqref{inf:etaeff_neg}], which is essentially the same case for the ultra-slow-roll (USR) inflation \cite{Kawasaki:1997ju, Choudhury:2013woa, Byrnes:2018txb, Carrilho:2019oqg, Cole:2022xqc}. Hence, the superhorizon growth of curvaton field perturbations in the non-minimal curvaton scenario also exists, which is confirmed by our numerical calculations for a concrete Gaussian-like field metric. Apart from a universal dip observed in the curvaton power spectrum, which arises from the cancellation between the constant mode and the ``decaying'' mode, a second dip exists due to the positively large effective friction term for the Gaussian-like field metric. With detailed theoretical and numerical analyses, we demonstrate that {\it the growth of curvaton field perturbations during inflation is not merely determined by the depth of the dip but also strongly affected by its first derivative}.

In addition, we employ the axion-like curvaton model, where the curvaton is treated as a pseudo-Nambu-Goldstone boson of a broken U(1) symmetry, exhibiting a periodic potential \cite{Dimopoulos:2003az} [cf. Eq.~\eqref{inf:potential}]. Reference \cite{Kasuya:2009up} for the first time found that the supersymmetry-based axion-like curvaton model can generate an extremely blue spectrum of isocurvature perturbations, in which the curvaton is identified with the phase direction of a complex scalar field. Subsequently, Ref. \cite{Kawasaki:2012wr} successfully realized the PBH formation based on this type of axion-like curvaton, and other works had utilized this model to account for LIGO/Virgo events \cite{Ando:2017veq, Ando:2018nge} or NANOGrav results \cite{Inomata:2020xad, Kawasaki:2021ycf}.

In the standard curvaton scenario \cite{Linde:1996gt, Lyth:2001nq, Lyth:2002my}, the curvaton is a spectator field $\chi$ that existed during inflation, which solely contributes to the primordial perturbations and has no effect on the inflationary background dynamics due to its negligible energy density ($\rho_\chi \ll \rho_\phi$) during inflation. After inflation, the inflaton completely decays into radiation when the Hubble parameter $H$ equals the decay rate of the inflaton $\Gamma_\phi$. 
The Hubble parameter then decreases, and the curvaton begins to oscillate when $H_\text{osc}$ is around $m_\chi$. During this oscillation phase, the curvaton behaves like dust, with its energy density scaling as $\bar{\rho}_\chi \propto a^{-3}$, becoming more dominant than radiation, with its energy density scaling as $\bar{\rho}_r \propto a^{-4}$, until the curvaton decays at a time $t_\text{dec}$ when $H = \Gamma_\chi$. 
For simplicity, we adopt the ``sudden-decay approximation'' such that the curvaton decays instantaneously into radiation\footnote{It has been turned out in Refs. \cite{Malik:2002jb, Malik:2006pm} that this approximation fits well to the full numerical result that includes the energy transfers in the whole post-inflation phase under the appropriate parameter choice.}. After $t_\text{dec}$, there is only radiation, and curvature perturbations thus remain conserved until it reenters the horizon.

Following the comprehensive analysis and formalism presented in Ref. \cite{Kawasaki:2011pd}, we consider the non-linear evolution of the curvaton field in the post-inflation dynamics, which has been demonstrated to have a significant impact on perturbation power spectrum and non-Gaussianity, consequently, the PBH abundance. We present complete expressions of the curvature power spectrum and non-Gaussianity at the curvaton's decay. Lastly, the PBH abundance and the SIGWs are calculated as well.

This paper is organized as follows. In Sec.~\ref{sec:inf}, we first review the full dynamics of the background and perturbations during inflation used for numerical calculations. Then, we identify a generic field metric as an effective friction term for curvaton field perturbations and find that the normal ``decaying'' mode starts growing on superhorizon scales when the total friction term becomes negative. The first derivative of field metric also plays an essential role in the growth of curvaton field perturbations during inflation. For a case study, we consider a Gaussian-like dip in the field metric, and our numerical results confirm the above conclusions. 
For the post-inflation dynamics, we apply the $\delta \mathcal{N}$ formalism to derive the analytic formalisms for the curvature power spectrum and the non-Gaussianity at the curvaton's decay. Numerical calculations are also performed. 
In Sec.~\ref{sec:pbh}, the PBH abundance and SIGW energy spectrum are derived numerically. 
We summarize the results in Sec.~\ref{sec:con}.

\section{Growth of perturbations in non-minimal curvaton scenario} \label{sec:inf}

\subsection{Inflationary dynamics}

\subsubsection{Master perturbation equations}

We consider the following action governing the dynamics of inflaton $\phi$ and curvaton $\chi$ during inflation,
\be \label{inf:action}
S = \int\ddd^4x \sqrt{-g} \l[ {M_{\mathrm{Pl}}^{2} \over 2} R - {1\over2} \nabla_\mu\phi\nabla^\mu\phi - {1\over2} \lambda^2(\phi) \nabla_\mu\chi\nabla^\mu\chi - V(\phi, \chi) \r]
\ee
with a non-trivial field metric $\lambda^2(\phi)$ that may arise in UV-completion theories \cite{Gomez-Reino:2006sqc,Covi:2008ea, Covi:2008cn}. Here $M_\text{Pl}$ is the reduced Planck mass, and $R$ is the Ricci scalar. 
For simplicity, we consider the minimal coupling between the inflaton and curvaton so that combined potential is separable, namely $V(\phi, \chi) = V_\text{inf}(\phi) + V_\text{cur}(\chi)$. The inflation's potential $V_\text{inf}(\phi)$ is required to be consistent with the standard slow-roll inflation, while the axion-like curvaton is written as
\be \label{inf:potential}
V_\text{cur}(\chi) = f_a^2 m_\chi^2 \l( 1 - \cos{\chi \over f_a} \r) ~,
\ee
where $m_\chi$ is the mass of $\chi$ and $f_a$ is its symmetry breaking scale. 
For the small-field limit $\chi \ll f_a$, the leading term in \eqref{inf:potential} gives the quadratic potential, namely $V_\text{cur}(\chi) \simeq {1\over2} m_\chi^2 \chi^2$, which is studied in Refs. \cite{Pi:2021dft, Meng:2022low}. 
The homogeneous EoMs derived from the action \eqref{inf:action} with the spatially flat Friedmann-Lema\^{\i}tre-Robertson-Walker metric, $\ddd s^2 = - \ddd t^2 + a^2(t) \delta_{ij} \ddd x^i \ddd x^j$, are given by
\bl \label{inf:hubble}
&3 M_{\mathrm{Pl}}^2 H^2 = \l[ {1\over2} \dot{\bar{\phi}}^2 + {\lambda^2(\bar{\phi}) \over 2} \dot{\bar{\chi}}^2 + V(\bar{\phi}, \bar{\chi}) \r] ~,
\\ \label{inf:phi_bg}
&\ddot{\bar{\phi}} + 3 H \dot{\bar{\phi}} + V_{,\phi}
= \lambda \lambda_{,\phi} \dot{\bar{\chi}}^2 ~,
\\ \label{inf:chi_bg}
&\ddot{\bar{\chi}} + \l( 3 H + 2 {\lambda_{,\phi} \over \lambda} \dot{\bar{\phi}} \r) \dot{\bar{\chi}} + {V_{,\chi} \over \lambda^2} = 0  ~,
\el
where $V_{,\chi} = f_a m_\chi^2 \sin{\chi \over f_a}$ for the axion-like potential \eqref{inf:potential} and the overhead bar represents the homogeneous background. 
Given the explicit form of field metric $\lambda(\phi)$ [cf. Eq. \eqref{inf:lambda}], the above three equations form a closed system and can be solved numerically.

The scalar fields can be separated into a homogeneous background and a first-order perturbation, namely $\phi(t,\textbf{x}) = \bar{\phi}(t) + \delta\phi(t,\textbf{x})$ and $\chi(t,\textbf{x}) = \bar{\chi}(t) + \delta\chi(t,\textbf{x})$. 
By incorporating metric perturbations, the dynamics of the field perturbations can be derived. The metric perturbation in the Newtonian gauge is written as,
\be\label{inf:metric_pert}
\ddd s^2 = -(1+2\Phi)\ddd t^2 + a(t)^2 \l[(1-2\Phi)\delta_{ij} + \frac{1}{2}h_{ij} \r] \ddd x^i \ddd x^j ~,
\ee
where the first-order scalar metric perturbations are described by a single variable $\Phi$, i.e., we ignore the anisotropic stress at the linear order, and $h_{ij}$ represents the second-order transverse-traceless tensor perturbations.
By employing the equations of first-order perturbation $\Phi$, we can express the field perturbation equations in a closed form \cite{Lalak:2007vi},  
\bl
\label{inf:eom_pert1}
&\ddot{Q}_\phi + 3H\dot{Q}_\phi - 2 \lambda \lambda_{,\phi} \dot{\bar{\chi}} \dot{Q}_\chi 
+ \left(\frac{k^2}{a^2}+C_{\phi\phi}\right)Q_\phi + C_{\phi\chi}Q_\chi =0 ~,
\\
\label{inf:eom_pert2}
&\ddot{Q}_\chi+3H\dot{Q}_\chi +2 {\lambda_{,\phi} \over \lambda} \dot{\bar{\phi}} \dot{Q}_\chi +2{\lambda_{,\phi} \over \lambda} \dot{\bar{\chi}} \dot{Q}_\phi 
+\left(\frac{k^2}{a^2}+C_{\chi\chi}\right)Q_\chi + C_{\chi\phi}Q_\phi =0 ~,
\el
where we introduce the gauge-invariant Mukhanov-Sasaki variables, $Q_\phi \equiv \delta\phi + {\dot{\bar{\phi}} \over H} \Phi$ and $Q_\chi \equiv \delta \chi + {\dot{\bar{\chi}} \over H} \Phi$, which correspond to $\delta\phi$ and $\delta\chi$ in the spatially-flat gauge $\Phi=0$. The background-dependent coefficients are defined as follows:
\bl
C_{\phi\phi} &= -2 (\lambda_{,\phi})^2 \dot{\bar{\chi}}^2 
+ {3\dot{\bar{\phi}}^2 \over M_\text{pl}^2}
- { \lambda^2 \dot{\bar{\phi}}^2 \dot{\bar{\chi}}^2 \over 2M_\text{pl}^4 H^2}
- {\dot{\bar{\phi}}^4 \over 2M_\text{pl}^4 H^2}
+ \left[ (\lambda_{,\phi})^2 - \lambda \lambda_{,\phi\phi} \right] \dot{\bar{\chi}}^2
+ {2\dot{\bar{\phi}} V_{,\phi} \over M_\text{pl}^2 H} + V_{,\phi\phi} ~,
\\
C_{\phi\chi} &= 
{3 \lambda^2 \dot{\bar{\phi}} \dot{\bar{\chi}} \over M_\text{pl}^2} - { \lambda^4 \dot{\bar{\phi}} \dot{\bar{\chi}}^3 \over 2M_\text{pl}^4 H^2} - { \lambda^2 \dot{\bar{\phi}}^3 \dot{\bar{\chi}} \over 2M_\text{pl}^4 H^2} +{ \dot{\bar{\phi}} V_{,\chi} \over M_\text{pl}^2 H} + {\lambda^2 \dot{\bar{\chi}} V_{,\phi} \over M_\text{pl}^2 H} +V_{,\phi\chi} ~,
\\
C_{\chi\chi} &=
{3 \lambda^2 \dot{\bar{\chi}}^2 \over M_\text{pl}^2} - {\lambda^4 \dot{\bar{\chi}}^4 \over 2M_\text{pl}^4 H^2} - { \lambda^2 \dot{\bar{\phi}}^2 \dot{\bar{\chi}}^2 \over 2M_\text{pl}^4 H^2} + {2\dot{\bar{\chi}} V_{,\chi} \over M_\text{pl}^2 H} + {V_{,\chi\chi} \over \lambda^2} ~,
\\
C_{\chi\phi} &=
{3\dot{\bar{\phi}} \dot{\bar{\chi}} \over M_\text{pl}^2} - { \lambda^2 \dot{\bar{\phi}} \dot{\bar{\chi}}^3 \over 2M_\text{pl}^4 H^2} - {\dot{\bar{\phi}}^3 \dot{\bar{\chi}} \over 2M_\text{pl}^4 H^2} 
+ 2 {\lambda_{,\phi\phi}\lambda  - (\lambda_{,\phi})^2 \over \lambda^2} \dot{\bar{\phi}} \dot{\bar{\chi}}
- 2 { \lambda_{,\phi} \over \lambda^3} V_{,\chi} 
 + {\dot{\bar{\phi}} V_{,\chi} \over M_\text{pl}^2 H \lambda^2}
 + {\dot{\bar{\chi}} V_{,\phi} \over M_\text{pl}^2H} + {V_{,\phi\chi} \over\lambda^2} ~, 
\el
The physical observables of interest are gauge-invariant curvature and isocurvature perturbations, defined as $\mathcal{R} \equiv \Phi + H {\dot{\bar{\phi}} \delta\phi + \dot{\bar{\chi}} \delta\chi \over \dot{\bar{\phi}}^2 + \dot{\bar{\chi}}^2}$ and $\mathcal{F} \equiv { -\dot{\bar{\chi}} \delta\phi + \dot{\bar{\phi}} \delta\chi \over \sqrt{ \dot{\bar{\phi}}^2 + \dot{\bar{\chi}}^2 } }$, respectively. For notational simplicity, we drop overhead bars of background quantities without ambiguity in the following discussions.

\subsubsection{Enhancement from a non-trivial field metric}

Although the dynamics of inflaton and curvaton perturbations during inflation are quite complicated [cf. Eqs. \eqref{inf:eom_pert1} and \eqref{inf:eom_pert2}], a reasonable estimate of curvaton evolution can be made \cite{Pi:2021dft}. 
During inflation, the curvaton is subdominant ($\rho_\chi \ll \rho_\phi$) and light ($m_\chi \ll H_\text{inf}$). As a result, $\chi$ is nearly frozen on its potential \eqref{inf:potential} due to the strong Hubble friction term.\footnote{Hence, the backreaction effect from $\chi$ on $\phi$ is negligible in the presence of the coupling $\lambda(\phi)$, see the right hand side of Eq.~\eqref{inf:phi_bg}.} Thus, we can neglect the terms involving $\dot{\chi}$ and temporarily ignore the coupling with the inflaton perturbation $\delta\phi$ in Eq. \eqref{inf:eom_pert2}, which gives us:
\be \label{inf:deltachi_apprx}
\delta\ddot{\chi}_k + \l(3 + 2 { \lambda_{,\phi} \over \lambda } {\dot{\phi} \over H} \r) H \delta\dot{\chi}_k + \l( \frac{k^2}{a^2} + {V_{,\chi\chi} \over \lambda^2} \r) \delta\chi_k \simeq 0 ~,
\ee
where ${V_{,\chi\chi} \over \lambda^2} = {m_\chi^2 \over \lambda^2} \cos{\chi \over f_a}$ for the axion-like potential \eqref{inf:potential}. The background \eqref{inf:chi_bg} and perturbation \eqref{inf:deltachi_apprx} equations do not share the same form as the case for the quadratic curvaton's potential \cite{Pi:2021dft, Meng:2022low}, and one cannot apply the fixed ratio $\delta\chi/\chi$ to simplify the analysis except for the small-field limit $\chi \ll f_a$.
Deep inside the horizon $k \gg a H$, the Bunch-Davies (BD) vacuum can be utilized to establish the initial condition for $\delta\chi_k$,
\be \label{inf:bd}
\delta\chi_k(t) \simeq { 1 \over \sqrt{2 k} a(t) \lambda(\phi(t)) } \exp\l(- ik \int {\ddd t\over a(t)} \r) ~.
\ee
At the horizon crossing $k=aH$, the power spectrum of $\delta\chi_k$ is simply approximated as
\be \label{inf:pchi}
\mathcal{P}_{\delta\chi}(k) \equiv {k^3 \over 2 \pi^2} |\delta\chi_k|^2 \simeq \l( {H_\text{inf} \over 2\pi \lambda(\phi)} \r)^2 \Bigg|_{k=aH} ~.
\ee
Naively, the dip in the field metric $\lambda(\phi)$ can enhance certain modes (around the scale that exits the horizon as $\phi$ passes the dip) of curvaton perturbations and lead to the large curvature spectrum after the curvaton's decay \cite{Pi:2021dft, Meng:2022low}. 
However, there are some discrepancies in the curvaton power spectrum at the end of inflation reported in Refs.~\cite{Pi:2021dft, Meng:2022low}, and the physical explanation of the enhancement is also missing. 
In what follows, we shall show that the enhancement of $\delta\chi$ arising from the non-trivial field metric $\lambda(\phi)$ has essentially the same physical reason with the case of the USR inflation \cite{Byrnes:2018txb,Carrilho:2019oqg}, and the estimate of curvaton power spectrum \eqref{inf:pchi} does not involve the superhorizon growth of curvaton perturbations.

The perturbation equation \eqref{inf:deltachi_apprx} clearly shows that the field metric $\lambda(\phi)$ plays a role as an effective friction term that is also well known in non-canonical inflation models for decades \cite{Armendariz-Picon:1999hyi, Armendariz-Picon:2000ulo, Domenech:2015qoa, DeAngelis:2023fdu, Lola:2020lvk, Choudhury:2023jlt, Choudhury:2023hvf}. Crucially, the effective friction term $2 { \lambda_{,\phi} \over \lambda } \dot{\phi}$ in Eq. \eqref{inf:deltachi_apprx} depends not only on the value of $\lambda$ but also on the first derivative $\lambda_{,\phi}$. It is physically transparent to define
\be \label{inf:etaeff}
\eta_\text{eff} \equiv - 2 \sqrt{2\epsilon} M_\text{Pl} { \lambda_{,\phi} \over \lambda } ~,
\ee
where $\epsilon \equiv -\dot{H}/H^2 = {1\over2M_\text{Pl}^2} (\dot{\phi}/H)^2 $ is the slow-roll parameter of inflationary background. One can rewrite Eq. \eqref{inf:deltachi_apprx} under the superhorizon limit $k\rightarrow0$ and ignore the effective mass term which is insignificant in our case,
\be \label{inf:usr}
\delta\ddot{\chi}_k + \l(3 + \eta_\text{eff} \r) H \delta\dot{\chi}_k + \frac{k^2}{a^2} \delta\chi_k \simeq 0 ~,
\ee
which is the same as the case for superhorizon comoving curvature perturbations, and there exists a constant mode and a time-evolving mode in its solution \cite{Byrnes:2018txb}, 
\be \label{inf:sol}
\delta\chi_k(t) \simeq C_k + D_k \int^t {\ddd \tilde{t} \over a^3(\tilde{t}) \epsilon_\text{eff}} ~,
\ee 
where $C_k$ and $D_k$ are functions only of $k$, and the effective slow-roll parameter $\epsilon_\text{eff}$ is normally defined and related to $\eta_\text{eff}$ as $\eta_\text{eff} = {\dot{\epsilon}_\text{eff} \over H \epsilon_\text{eff}}$. 
One can clearly see the physical relevance between Eq. \eqref{inf:usr} and USR inflation \cite{Byrnes:2018txb, Carrilho:2019oqg}. The total friction term $(3+\eta_\text{eff})H$ is negative when
\be \label{inf:etaeff_neg}
\eta_\text{eff} < -3 ~,
\ee
and the second term in Eq.~\eqref{inf:sol} will become a growing mode and can dominate the solution at a later time. The curvaton perturbation will grow on superhorizon scales and enhance the curvaton perturbation power spectrum at the end of inflation. 
Hence, the curvaton spectrum at the end of inflation must differ from Eq. \eqref{inf:pchi}, which is only valid if the curvaton perturbations were frozen after the horizon exit. Moreover, the canonicalized curvaton perturbation is not simply $\lambda(\phi)\delta\chi$ (which is also a hidden assumption of Eq. \eqref{inf:pchi}) especially for the enhanced modes, due to the corresponding non-negligible derivatives $\lambda_{,\phi}$, see the right panel of Fig.~\ref{fig:lambda}. {\it In summary, the growth of curvaton perturbations during inflation is not solely determined by the value $\lambda$ but also strongly affected by the first derivative $\lambda_{,\phi}$).}

\subsubsection{Case study: the Gaussian-like dip}

\begin{figure}[ht]
	\centering
	\includegraphics[width=0.32\textheight]{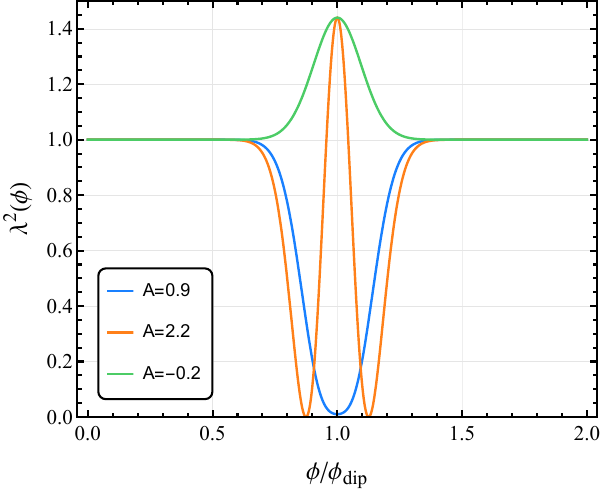}
	\includegraphics[width=0.33\textheight]{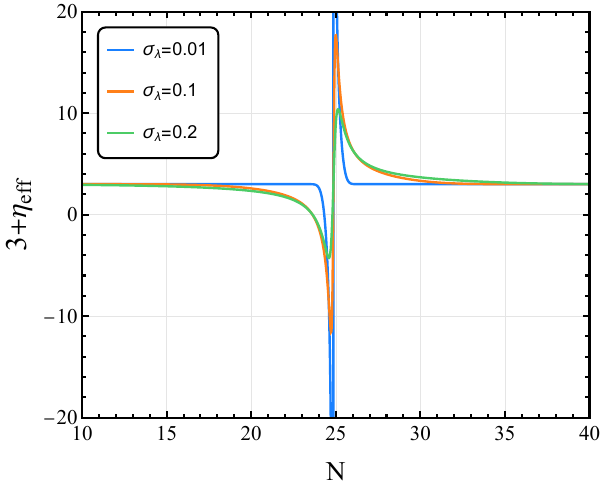}
	\caption{{\it Left panel}: An illustration of the field metric $\lambda^2(\phi)$ in terms of three typical values of $A$: $0.9$ (blue), $2.2$ (orange) and $-0.2$ (green), with the same width $\sigma_\lambda = 0.1$. This work only focuses on the first case $0 < A < 1$. {\it Right panel}: The numerical results of $3 + \eta_\text{eff}$ during the Starobinsky inflation for $\sigma_\lambda = (0.01, 0.1, 0.2)$ shown by the blue, orange and green curves, respectively. The parameters are taken as: $m_\chi/M_\text{Pl} = 10^{-8}$, $\phi_\text{ini}/M_\text{Pl} = 5.5$, $\phi_\text{dip}/M_\text{Pl} = 4.8$, $A=0.995$, $\sigma_\lambda = 0.01$ and $\Lambda^4/M_\text{Pl}^4 \simeq 2 \times 10^{-14}$.}
	\label{fig:lambda}
\end{figure}

To further elaborate on the aforementioned points, we shall examine a Gaussian-like dip in $\lambda(\phi)$ as Ref.~\cite{Meng:2022low},
\be \label{inf:lambda}
\lambda(\phi) = \lambda_c \l\{ 1 - A \exp{\l[ - {(\phi - \phi_\text{dip})^2 \over 2 \sigma_\lambda^2} \r]} \r\} ~,
\ee
where the overall amplitude is governed by $\lambda_c = 1$, the depth of the dip at $\phi_\text{dip}$ is determined by $A$ and the width is controlled by $\sigma_\lambda$.
The choice of the value of $\lambda_c$, which is close to the canonical case ($A=0$), is based on the premise that we are solely concerned with the physical impact generated by its dip, namely the enhancement of $\delta\chi$ near $\phi_\text{dip}$ as we will show later.
It is worth noting that when $A = 1$, the kinetic term of $\chi$ in the action \eqref{inf:action} vanishes at $\phi_\text{dip}$, resulting in the deactivation of $\chi$'s dynamics. Hence, $\chi$ swiftly approaches the ground state $\chi = 0$ and remains there permanently. Consequently, the standard curvaton scenario is absent in the post-inflation period, which is not our interest in this paper. 
For $A > 1$, two zero points and one local maximum exist in $\lambda^2(\phi)$. However, the curvaton will be trapped as it encounters the first dip.
Moreover, if $A$ is negative, then the global minimum of $\lambda^2(\phi)$ is unity, as illustrated in the left panel of Fig.~\ref{fig:lambda}. In order to consider the growth of curvaton perturbation $\delta\chi$, we will focus on the range $0 < A < 1$ in the subsequent discussion.

\begin{figure}[ht]
	\centering
	\includegraphics[width=0.35\textheight]{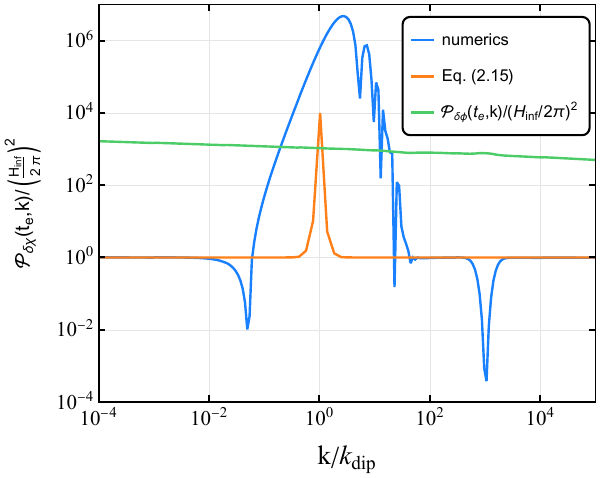}
	\includegraphics[width=0.32\textheight]{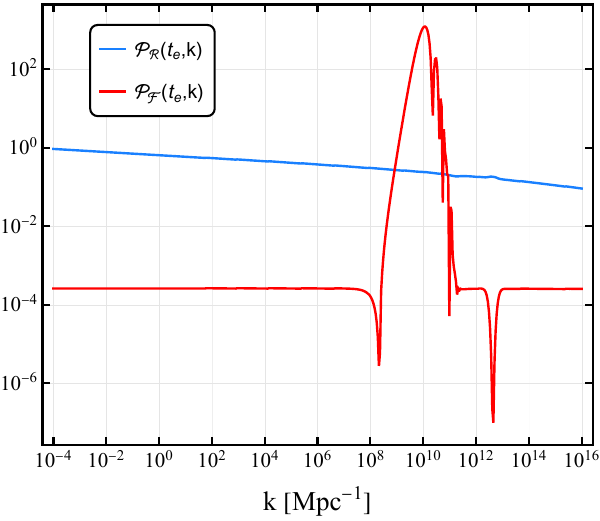}
	\caption{ {\it Left panel}: The comparison between our numerical result (the blue curve) and the approximation \eqref{inf:pchi} adopted by Ref. \cite{Pi:2021dft} (the orange dashed curve), for the curvaton power spectrum $\mathcal{P}_{\delta\chi}(t_\text{e},k)$. The inflaton power spectrum $\mathcal{P}_{\delta\phi}(t_\text{e},k)$ is shown by the green curve.
	{\it Right panel}: The numerical results of the comoving curvature spectrum $\mathcal{P}_{\mathcal{R}}(t_\text{e},k)$ (blue) and entropy spectrum $\mathcal{P}_{\mathcal{F}}(t_\text{e},k)$ (red) at the end of inflation, which are normalized by $\mathcal{P}_{\mathcal{R}}(t_\text{e},k_\text{pivot})$.
	The parameters are taken as: $m_\chi/M_\text{Pl} = 10^{-8}$, $\phi_\text{ini}/M_\text{Pl} = 5.5$, $\phi_\text{dip}/M_\text{Pl} = 4.8$, $A=0.995$, $\sigma_\lambda = 0.01$ and $\Lambda^4/M_\text{Pl}^4 \simeq 2 \times 10^{-14}$.}
	\label{fig:pchi}
\end{figure}

\begin{figure}[ht!]
	\centering
	\includegraphics[width=0.33\textheight]{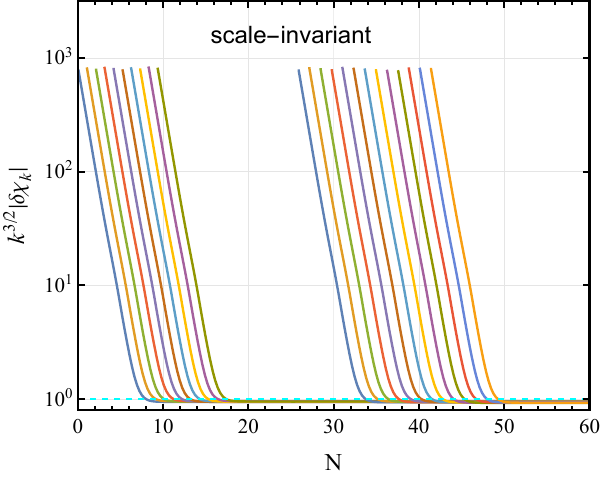}
	\includegraphics[width=0.34\textheight]{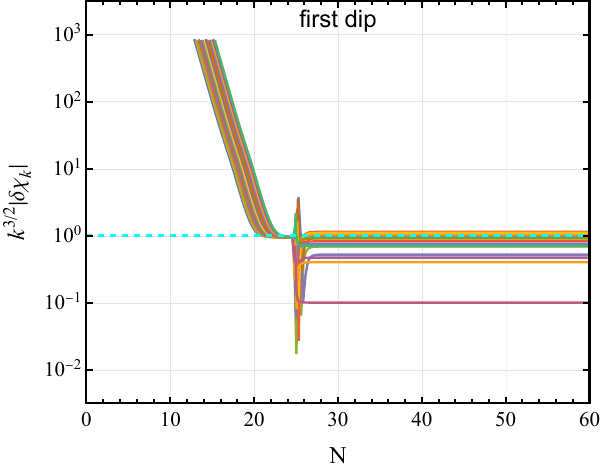}
	\includegraphics[width=0.34\textheight]{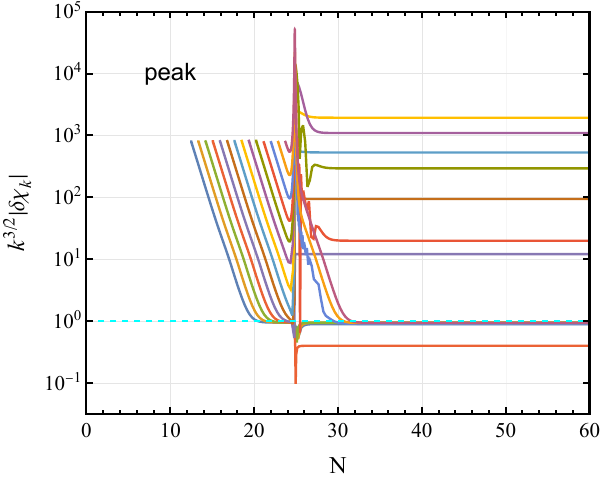}
	\includegraphics[width=0.34\textheight]{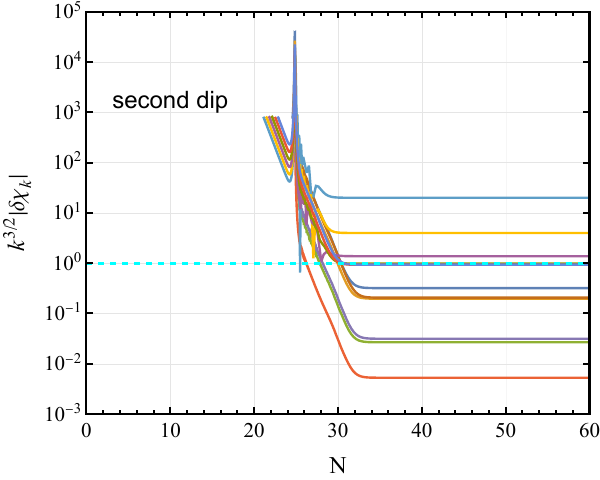}
	\includegraphics[width=0.34\textheight]{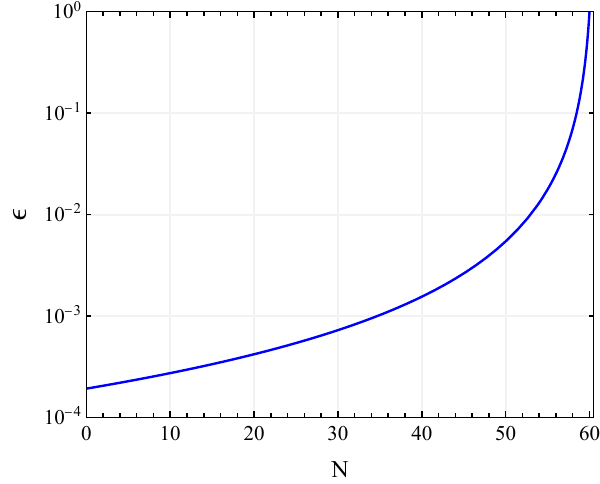}
	\includegraphics[width=0.34\textheight]{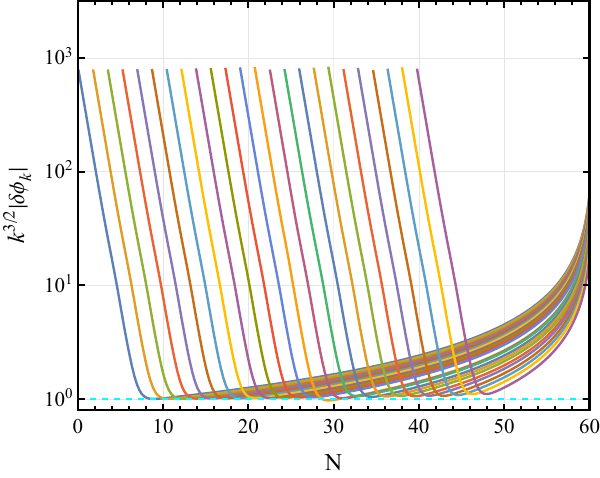}
	\caption{The numerical results of time evolutions of $|\delta\chi_k|$ and $|\delta\phi_k|$ during the 60-efolding Starobinsky inflation with the same parameters as Fig. \ref{fig:pchi}. In each panel, later start times correspond to larger $k$ modes. The upper four panels display different $k$ ranges corresponding to the scale-invariant, first/second dip and peak regimes of $\mathcal{P}_{\delta\chi}(t_\text{e},k)$ as shown in Fig. \ref{fig:pchi}. The bottom-right panel clearly shows the superhorizon growth of $|\delta\phi_k|$ in the presence of an increasing slow-roll parameter $\epsilon$ of Starobinsky inflation shown in the bottom-left panel. 
	The cyan dashed horizontal line refers to the value $H_\text{inf}/\sqrt{2}$. 
}
	\label{fig:time_evol}
\end{figure}

After numerically solving the master perturbation equations \eqref{inf:eom_pert1} and \eqref{inf:eom_pert2} in the Starobinsky inflation $V_\text{inf}(\phi) = \Lambda^4 \l[ 1 - \exp\l( {- \sqrt{{2\over3}} {\phi \over M_\text{Pl}} } \r)\r]^2$, which has been demonstrated to be in good agreement with Planck's observations \cite{Planck:2018jri}, we yield the power spectra for inflaton (green) and curvaton (blue) at the end of inflation\footnote{For the purpose of numerical calculations, the moment at which the slow-roll parameter $\epsilon$ reaches unity is considered as the end of inflation.} for the axion-like potential \eqref{inf:potential}, as shown in the left panel in Fig.~\ref{fig:pchi}, respectively\footnote{Note that the energy scale $\Lambda^4$ of Starobinsky inflation in our case is lower than the typical value $\sim 10^{-10}$ in the single-field case \cite{Planck:2018jri}, since there exits additional contribution from $\delta\chi$ in the post-inflation epoch, as we will discuss later.}. Here, we set the initial inflaton field value as $\phi_\text{ini}/M_\text{Pl} = 5.5$. Reference \cite{Meng:2022low} also reported a similar curvaton spectrum as our numerical result, which is distinct from the approximation \eqref{inf:pchi} (orange) adopted by Ref. \cite{Pi:2021dft}.
The peak of $\mathcal{P}_{\delta\chi}(t_\text{e},k)$ predicted by Eq. \eqref{inf:pchi} is about three orders of magnitude smaller than ours and much narrower as well. As discussed previously, the omission of the enhancement effect arising from $\lambda_{,\phi}$ is the cause for these significant deviations. 
The numerical results for the comoving curvature spectrum $\mathcal{P}_{\mathcal{R}}(t_\text{e},k)$ (blue) and entropy spectrum $\mathcal{P}_{\mathcal{F}}(t_\text{e},k)$ (red) are shown in the right panel of Fig. \ref{fig:pchi}. Since the curvaton is assumed to be light during inflation, the inflationary trajectory is along the $\phi$-direction. The curvature and entropy perturbations are decoupled 
and dominated by $\delta\phi$ and $\delta\chi$, respectively. Hence, both $\mathcal{P}_{\mathcal{R}}(t_\text{e},k)$ and $\mathcal{P}_{\mathcal{F}}(t_\text{e},k)$ share the similar shapes with $\mathcal{P}_{\delta\phi}(t_\text{e},k)$ and $\mathcal{P}_{\delta\chi}(t_\text{e},k)$, respectively. The slow-roll parameter $\epsilon$ increases when $\phi$ approaches the end of inflation, as shown in the bottom-left panel of Fig. \ref{fig:time_evol}, $\delta\phi$ is thus not frozen on superhorizon scales and causes $\mathcal{P}_{\delta\phi}(t_\text{e},k)$ to be larger than (around three orders of magnitude) the standard amplitude $(H_\text{inf}/2\pi)^2$ for a light scalar field during inflation\footnote{The value of Hubble parameter $H_\text{inf} \simeq \Lambda^2$ in our numerical calculation is taken to normalize the spectrum $\mathcal{P}_{\delta\chi}(t_\text{e},k)$ to be $(H_\text{inf}/2\pi)^2$ on large scales, and the benchmark $\mathcal{P}_{\zeta}(t_\text{dec},k_\text{pivot}) \simeq 2.1 \times 10^{-9}$ in Eq. \eqref{post:pzeta} as well.}, as displayed in the left panel of Fig. \ref{fig:pchi} and the bottom-right panel of Fig. \ref{fig:time_evol}.

The left panel of Fig. \ref{fig:pchi} displays three manifest features of $\mathcal{P}_{\delta\chi}(t_\text{e}, k)$: two dips and one peak in between.
First, a pronounced dip appears prior to the growth of $\mathcal{P}_{\delta\chi}(t_\text{e}, k)$, which is essentially the same as USR inflation as discussed earlier, such that the decaying mode cancels with the constant mode in the solution \eqref{inf:sol} to some extent\footnote{These two modes can cancel with each other exactly for USR inflation ($\eta_\text{eff}=-6$), see Refs.~\cite{Byrnes:2018txb, Carrilho:2019oqg} for the detailed discussions.}, signalling the superhorizon growth of this ``decaying mode'' (which is a growing mode now) and results in the following growth of $\mathcal{P}_{\delta\chi}(t_\text{e}, k)$. The spectral tilt of growth is around $k^4$ that satisfies the bound discussed in Refs.~\cite{Byrnes:2018txb,Carrilho:2019oqg}, since the non-minimal curvaton model is essentially the single-field inflation. For small-$k$ modes shown by the first group of $k$s in the top-left panel of Fig. \ref{fig:time_evol}, they exit the horizon at earlier times (i.e., around the turning points displayed in this plot), and the decaying mode has already decreased to an extremely small value before the total friction term becomes negative, so that the constant mode always dominate $\delta\chi$ even though the ``decaying mode'' constantly grows during $\eta_\text{eff} < -3$ on superhorizon scales. Thus, one can easily realize the nearly scale-invariant spectrum on large scales as long as the inflaton starts at a position (namely the pivot scale $k_\text{pivot}$ exits the horizon) not too close to the dip position $\phi_\text{dip}$. It is evident in the top-left panel of Fig. \ref{fig:time_evol} that these scale-invariant modes behave the same as BD vacuum modes, which decay as $|\delta\chi_k(t)| \sim 1/a(t)$ on subhorizon scales revealed by Eq. \eqref{inf:bd}.

Second, the growth of curvaton perturbation occurs within a certain $k$ range, which is caused by the negative total friction term in Eq. \eqref{inf:usr}, namely $\eta_\text{eff} < -3$, when the inflaton approaches the dip position $\phi_\text{dip}$ from the right side in the left panel of Fig.~\ref{fig:lambda}. And $\eta_\text{eff}$ is negative only when $\phi > \phi_\text{dip}$ for the Gaussian dip \eqref{inf:lambda}. The $k$ range of the enhancement approximately corresponds to the horizon-crossing modes when $\eta_\text{eff} < -3$, illustrated in the middle-left panel of Fig.~\ref{fig:time_evol}, whose final values (at the e-folding number $N=60$) are higher than the scale-invariant modes (the cyan dashed line) shown in the top-left panel of Fig. \ref{fig:time_evol}.
Moreover, the spectrum $\mathcal{P}_{\delta\chi}(t_\text{e},k)$ drops after it reaches the crown since the total friction returns to zero and then becomes positively large as shown in the right panel of Fig.~\ref{fig:lambda}.

Last but not least, there exists a second dip in the spectrum $\mathcal{P}_{\delta\chi}(t_\text{e}, k)$ after it stops falling, since the total friction term becomes positively large (see the right panel of Fig.~\ref{fig:lambda}) when the inflaton passes over the dip position $\phi_\text{dip}$. It is evident in the middle-right panel of Fig. \ref{fig:time_evol} that certain modes $|\delta\chi_k|$ drop more rapidly after their growth, the slopes of which are larger than the normal BD vacuum modes $|\delta\chi_k(t)| \sim 1/a(t)$. Reference \cite{Meng:2022low} reported the first dip without explanation and did not mention the presence of the second dip, as we show here. The first and second dips may be weakened in the final curvature power spectrum since it will be compensated by the nearly scale-invariant $\mathcal{P}_{\delta\phi}(t_\text{e},k)$, see Eq. \eqref{post:pzeta} and the curves shown in Fig. \ref{fig:totpzeta}.

Moreover, it is necessary to clarify the parameter dependence of the features mentioned above, and we will see that the first derivative $\lambda_{,\phi}$ plays an essential role in each case.
Examining the axion-like potential \eqref{inf:potential} and the Gaussian dip \eqref{inf:lambda}, the free parameters in our model are given by $(m_\chi, f_a, A, \phi_\text{dip}, \sigma_\lambda)$\footnote{These free parameters receive constraints from the curvature power spectrum as shown in Fig.~\ref{fig:totpzeta}.}. In the following discussions, we fix $f_a$, which is somewhat irrelevant to the inflationary dynamics since it only changes the normalization of $\chi$ (which is nearly frozen during inflation due to the smallness of $m_\chi$)\footnote{However, $f_a$ is crucial for the enhancement of the final curvature perturbation at the curvaton's decay, since the first derivative $N_{,\chi_\text{e}}$ is inversely proportional to $f_a$ or $\chi_\text{e}$, see Eqs. \eqref{post:deriv1} and \eqref{post:pzeta}.} and can be absorbed into its mass $m_\chi$ in the axion-like potential \eqref{inf:potential}. The dip position $\phi_\text{dip}$ determines the peak position of the curvaton perturbation spectrum $\mathcal{P}_{\delta\chi}(t_\text{e}, k)$ (or equivalently, the peak of the final curvature spectrum $\mathcal{P}_{\zeta}(t_\text{dec}, k)$  shown in Fig. \ref{fig:totpzeta}), so that we mainly focus on the discussion about $A$ and $\sigma_\lambda$ for the moment.

\begin{figure}[ht!]
	\includegraphics[width=0.35\textheight]{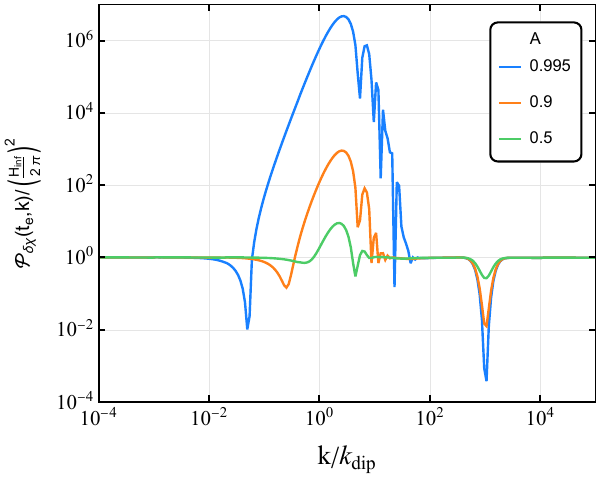}
	\includegraphics[width=0.35\textheight]{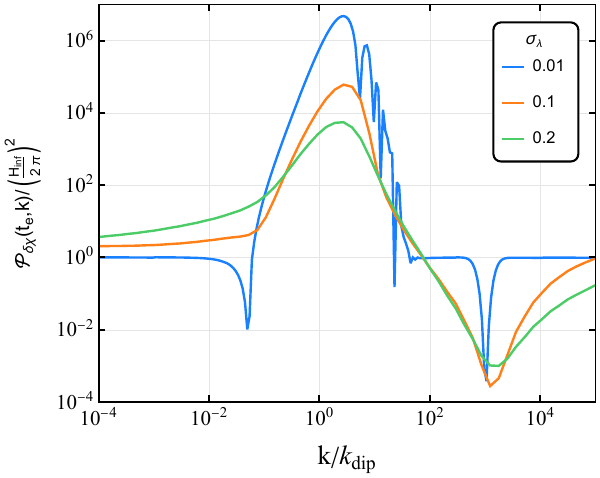}
	\caption{{\it Left panel}: The curvaton power spectra $\mathcal{P}_{\delta\chi}(t_\text{e}, k)$ in terms of various values $A = (0.995, 0.9, 0.5)$ for a fixed value $\sigma_\lambda = 0.01$.
	{\it Right panel}: The curvaton power spectra $\mathcal{P}_{\delta\chi}(t_\text{e}, k)$ in terms of various values $\sigma_\lambda = (0.01, 0.1, 0.2)$ for a fixed value $A = 0.995$. All other parameters are the same as Fig. \ref{fig:pchi}.}
	\label{fig:A}
\end{figure}

For a fixed $\sigma_\lambda$, a larger $A$ refers to a dip with a steeper slope, or equivalently, $\eta_\text{eff}$ defined in Eq.~\eqref{inf:etaeff} is enlarged, which results in a more significant enhancement. This trend can be identified from the comparisons among cases $A = (0.995, 0.9, 0.5)$ as illustrated in the left panel of Fig.~\ref{fig:A}. These peak amplitudes are not simply proportional to $[ \lambda(\phi_\text{dip}) ]^{-2} = (1-A)^{-2}$ as suggested by the approximation \eqref{inf:pchi}, which demonstrates again that the enhancement is also affected by $\lambda_{,\phi}$. For a smaller $A$ case, the peak and two dips become insignificant, as expected.

For a fixed $A$, an enlarged $\sigma_\lambda$ refers to a dip with a more gentle slope, the enhancement effect from $\lambda_{,\phi}$ becomes weaker. The maximum of the total friction is less than that of a smaller $\sigma_\lambda$ (see the evolution of the total friction as shown in the right panel of Fig. \ref{fig:lambda}), which therefore leads to a less enhancement of peak displayed in the right panel of Fig. \ref{fig:A}. 
For a wide dip with $\sigma_\lambda = 0.2$, the peak amplitude of $\mathcal{P}_{\delta\chi}(t_\text{e},k)$ is comparable with the approximation \eqref{inf:pchi} shown in the left panel of Fig. \ref{fig:pchi}. However, the shapes of $\mathcal{P}_{\delta\chi}(t_\text{e},k)$ are still distinct, due to the effect arising from $\lambda_{,\phi}$.
We notice that the first dip prior to the growth becomes shallower ($\sigma_\lambda = 0.1$) or even disappears for a wider dip ($\sigma_\lambda = 0.2$). This trend results from the fact that the condition \eqref{inf:etaeff_neg} occurs at an earlier time for a larger $\sigma_\lambda$ (see the right panel of Fig. \ref{fig:lambda}), which results in the growth of ``decaying mode'' in $\delta\chi$. It dominates over the constant mode at an earlier time (corresponding to a smaller $k$). This fact can also be observed from the slopes of $\mathcal{P}_{\delta\chi}(t_\text{e},k)$ in the small-$k$ region for $\sigma_\lambda = 0.1$ and $\sigma_\lambda = 0.2$, such that the growth of $\mathcal{P}_{\delta\chi}(t_\text{e},k)$ becomes earlier. 
Additionally, both cases display a broader second dip, which is caused by the fact that the positively large total friction term with a larger $\sigma_\lambda$ lasts for a longer time, as shown in the right panel of Fig. \ref{fig:lambda}.

\subsection{Post-inflation dynamics}

In what follows, we shall examine the dynamics during the post-inflation phase, namely from the end of inflation $H = \Gamma_\phi$ to the curvaton's decay $H = \Gamma_\chi$. Here, we adopt the sudden-decay approximation for simplicity, such that the energy transfer between the radiation and curvaton is negligible until $H = \Gamma_\chi$, such that $\chi$ decays to radiation completely and instantly (on the total uniform density hypersurface). To systematically deal with the non-linear evolutions of curvaton perturbation $\delta\chi$, we apply the $\delta \mathcal{N}$ formalism \cite{Starobinsky:1985ibc, Sasaki:1995aw, Lyth:2004gb, Abolhasani:2019cqw} to analytical estimates and numerical calculations of the superhorizon curvature perturbations in the post-inflation phase.

\subsubsection{$\delta \mathcal{N}$ formalism}

The $\delta \mathcal{N}$ formalism is a powerful tool that enables us to analytically calculate the nonlinear evolutions of scalar perturbations on large scales by solely solving the background equations without any knowledge of complicated perturbation dynamics. 
This formalism establishes a connection between the superhorzion curvature perturbation on the uniform-density hypersurface and the perturbation in e-folding number between the initial spatially-flat hypersurface $\Sigma_s(t_i)$ and the final uniform-density hypersurface $\Sigma_u(t_f)$, namely
\be \label{post:zeta_dn}
\zeta(t_f, \mathbf{x}) = N(t_i, t_f; \mathbf{x}) - \bar{N}(t_i, t_f) ~,
\ee
where $\bar{N}$ measures the homogeneous expansion from $\Sigma_s(t_i)$ to $\Sigma_s(t_f)$, while $N$ refers to the local expansion from $\Sigma_s(t_i)$ to $\Sigma_u(t_f)$. The non-linear definition of $\zeta$ without imposing $\delta\rho=0$ is given by \cite{Lyth:2004gb}, $\zeta \equiv - \psi + \frac13 \int_{\bar{\rho}(t)}^{\rho(t,\mathbf{x})} \frac{\ddd \tilde{\rho}}{ \tilde{\rho} + P(\tilde{\rho})}$ along with the parametrization of the full spatial metric $g_{ij}(t,\mathbf{x}) = a^2(t) e^{2 \psi(t,\mathbf{x})} \gamma_{ij}$, where $\psi$ is the non-linear generalization of the linear metric perturbation $\Phi$ shown in Eq. \eqref{inf:metric_pert}. In the $\delta \mathcal{N}$ formalism \eqref{post:zeta_dn}, the final time $t_f$ should be chosen as a moment such that our Universe has already reached the adiabatic limit, i.e., $\zeta$ is time-independent afterward. Hence, we choose $t_f$ as the curvaton's decay since $\zeta$ is conserved after $t_\text{dec}$ with a single matter component (radiation) existing in the Universe. For the initial time $t_i$, we choose it to be the end of inflation $t_\text{e}$ instead of the horizon-crossing usually adopted in other literature (e.g., \cite{Sasaki:2006kq, Kawasaki:2011pd, Ichikawa:2008iq, Kawasaki:2012wr, Kawasaki:2021ycf, Kobayashi:2020xhm}), since all the scales of interest are well outside the horizon at $t_\text{e}$ and the curvaton field $\chi$ has the non-trivial evolution between the horizon-crossing to the end of inflation for non-quadratic potentials \cite{Kawasaki:2011pd}. The underlying reason that one can choose the initial time to start the $\delta \mathcal{N}$ calculation ``at will'' is due to the intriguing property of the $\delta \mathcal{N}$ formalism \eqref{post:zeta_dn}, such that it is independent of the initial hypersurface. More importantly, the $\delta \mathcal{N}$ formalism states that one can write the Taylor expansion of the final superhorizon curvature perturbation on $\Sigma_s(t_i)$ as
\be \label{post:deltaN}
\zeta(\mathbf{x}) = N_{,\phi^a} \delta\phi^a + {1\over2} N_{,\phi^a\phi^b} \l[ \delta\phi^a \delta\phi^b - \langle \delta\phi^a\delta\phi^b \rangle \r] + \mathcal{O}(\delta\phi^3) ~,
\ee
where the summation is implied with repeated indices of $\phi^a$.
For notational simplicity, we drop the time arguments on both sides, and one should keep in mind that all the derivatives and field perturbations are evaluated on the initial hypersurface $\Sigma_s(t_i)$. 
On the other hand, the local non-Gaussianity of curvature perturbations is described in the following form,
\be \label{post:local}
\zeta(\mathbf{x}) = \zeta_g(\mathbf{x}) + {3\over5}f_\text{NL} \l[ \zeta_g^2(\mathbf{x}) - \langle \zeta_g^2(\mathbf{x}) \rangle \r] + \mathcal{O}(\zeta_g^3) ~,
\ee
where $\zeta_g(\mathbf{x})$ is the Gaussian part of $\zeta(\mathbf{x})$ satisfying the ensemble average $\langle \zeta_g \rangle = 0$.
Hence, matching Eq. \eqref{post:local} with Eq. \eqref{post:deltaN}, one immediately obtains the following relations,
\bl \label{post:pzeta_dn}
\mathcal{P}_\zeta &= \l( N_{,\phi}\r)^2 \mathcal{P}_{\delta\phi} + \l(N_{,\chi}\r)^2 \mathcal{P}_{\delta\chi}  ~,
\\ \label{post:fnl_dn}
f_\text{NL} &= {5\over6} { \l(N_{,\phi}\r)^2 N_{,\phi\phi} + \l(N_{,\chi}\r)^2 N_{,\chi\chi} \over \l[ \l(N_{,\phi}\r)^2 + \l(N_{,\chi}\r)^2 \r]^2} ~.
\el
In the above steps, $\delta\phi$ and $\delta\chi$ are assumed to be uncorrelated random fields at the time of interest. 
In the following calculations, we also assume that the curvaton field perturbation $\delta\chi_\text{e} \equiv \delta\chi(t_\text{e})$ is Gaussian, which is reasonable since $\chi$ is assumed to be a light field during inflation, and its field perturbation $\delta\chi$ should be scale-invariant and Gaussian during inflation \cite{Maldacena:2002vr}.

\begin{figure}[ht]
	\centering
	\includegraphics[width=0.32\textheight]{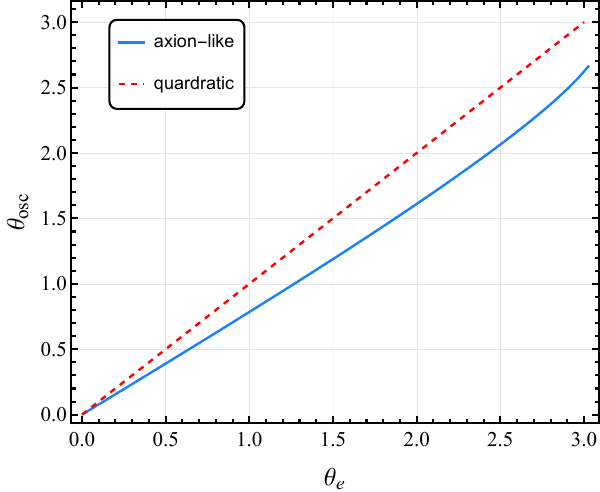}
	\includegraphics[width=0.32\textheight]{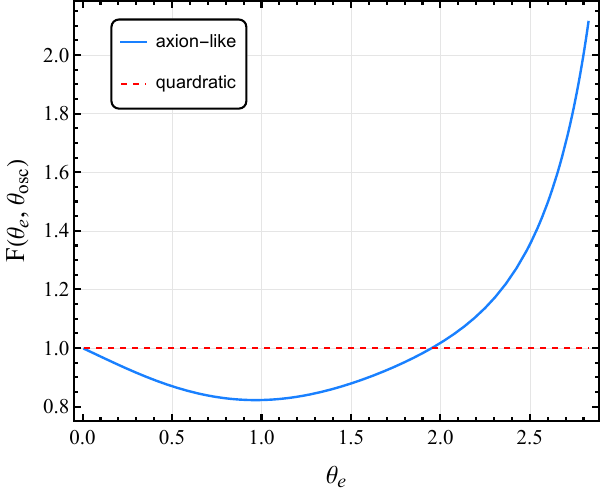}
	\caption{{\it Left panel}: The curvaton field value $\theta_\text{osc}$ as a function of $\theta_\text{e}$ for the axion-like potential (the solid blue curve) and for the quadratic potential (the red dashed curve), respectively.
	{\it Right panel}: The function $F(\theta_\text{e}, \theta_\text{osc})$ (the solid blue curve), defined in Eq. \eqref{post:ffunc}, describes the deviation of curvaton power spectrum from the quadratic potential (the red dashed line) in terms of the non-linear evolution of $\chi$ shown in the left panel.}
	\label{fig:nonelov}
\end{figure}

\begin{figure}[ht]
	\centering
	\includegraphics[width=0.32\textheight]{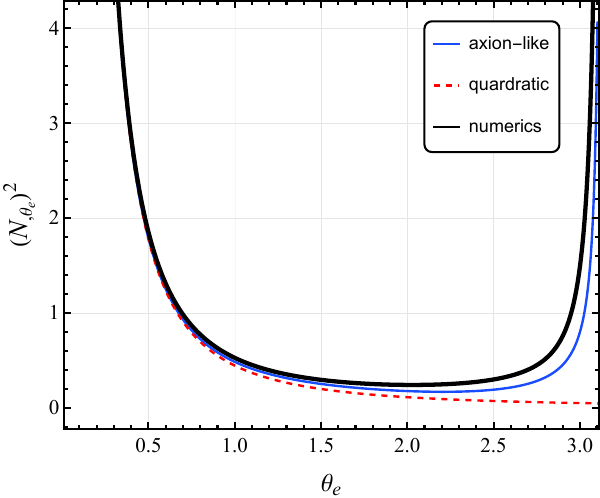}
	\includegraphics[width=0.32\textheight]{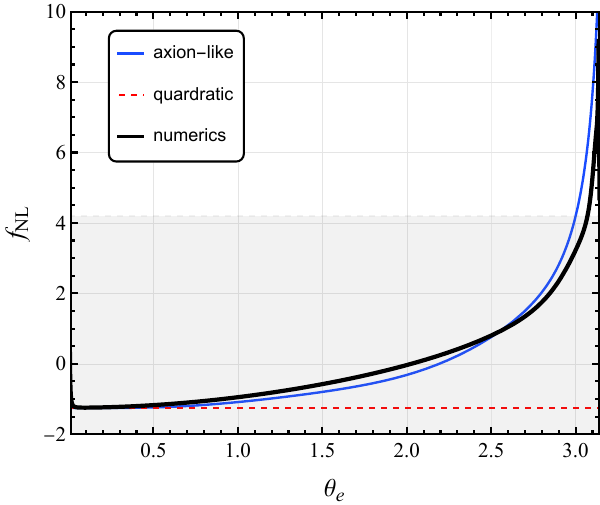}
	\caption{{\it Left panel}: The first derivative of e-folding number $(N_{,\theta_\text{e}})^2$ as a function of $\theta_\text{e}$.
		{\it Right panel}: The non-linear parameter $f_\text{NL}$ in terms of $\theta_\text{e}$ and the shadowed region arises from the Planck constraint $f_\text{NL} = -0.9 \pm 5.1$ \cite{Planck:2019kim}. In both panels, The black/blue solid and red dashed curves represent our numerical results, analytic formulas based on \eqref{post:deriv1} and \eqref{post:deriv2} and analytic formulas for the quadratic potential, respectively. }
	\label{fig:fnl}
\end{figure}

In what follows, we shall apply the formalism developed in Ref. \cite{Kawasaki:2011pd} to our mixed scenario, which is suitable for a general curvaton's potential. The most crucial conclusion presented in Ref. \cite{Kawasaki:2011pd} is that the statistical properties of curvature perturbations receive the potentially significant correction from the non-uniform onset of curvaton's oscillation, that typically arises in non-quadratic potentials. 
The onset of the curvaton's oscillation is naturally defined as a moment when the time scale of the curvaton rolling becomes comparable to the Hubble time \cite{Kawasaki:2011pd}, namely $H_\text{osc} \equiv \l| \dot{\chi} / \chi \r|$, instead of the usual uniform one $H_\text{osc} = m_\chi$ for a quadratic potential. 
In the period $(t_\text{e}, t_\text{osc})$, the radiation is still dominant, and the Hubble friction term is much stronger than the curvaton's mass term, so that one can use the slow-roll approximation to the curvaton's evolution in this period \cite{Kawasaki:2011pd},
\be \label{post:chi_evol}
5 H \dot{\chi} \simeq - V_{,\chi} ~,
\ee
which is subject to the condition $|V_{,\chi\chi} / 5 H^2|\ll1$ that holds during the period $(t_\text{e}, t_\text{osc})$ and $\dot{H}/H^2 = -2$ during the radiation-dominated era. With above preparations, the Hubble parameter at $t_\text{osc}$ is identified as a unique function of $\chi_\text{osc}$,
\be \label{post:hosc}
H_\text{osc}^2 \simeq { V_{,\chi}(\chi_\text{osc}) \over 5 \chi_\text{osc} }
= {1\over 5 \chi_\text{osc}} f_a m_\chi^2 \sin\l( {\chi_\text{osc} \over f_a} \r) ~,
\ee
which is consistent with $H_\text{osc}^2 \simeq m_\chi^2$ in the small-field limit $\chi_\text{osc} \ll f_a$ for the axion-like potential \eqref{inf:potential}. Integrating Eq. \eqref{post:chi_evol}, one can obtain the non-linear evolution of the curvaton field in the period $(t_\text{e}, t_\text{osc})$,
\be \label{post:chi_non}
{ \partial\chi_\text{osc} \over \partial\chi_\text{e} } = {1 \over 1 - Y(\chi_\text{osc})} { V_{,\chi}(\chi_\text{osc}) \over V_{,\chi}(\chi_\text{e}) } ~,
\ee
where $Y(\chi_\text{osc}) \equiv {1\over4} \l( {\chi_\text{osc} V_{,\chi\chi}(\chi_\text{osc}) \over V_{,\chi}(\chi_\text{osc})} - 1\r) = {\chi_\text{osc} \cot(\chi_\text{osc}/ f_a) - f_a \over 4 f_a}$, and we solve Eq. \eqref{post:chi_non} as
\be
\ln \l| { \tan(\theta_\text{osc}/2) \over \tan(\theta_\text{e}/2) } \r|
= - {\theta_\text{osc} \over 4 \sin\theta_\text{osc} } ~,
\ee
which is shown by the solid blue curve in the left panel of Fig. \ref{fig:nonelov}, and its asymptotic behavior can be written as
\be \label{post:asymp}
\theta_\text{osc}
\simeq
\theta_\text{e} - 0.286 \theta_\text{e}^2 + 0.086 \theta_\text{e}^4 - 0.012 \theta_\text{e}^6 + 0.0006 \theta_\text{e}^8 ~,
\ee
where $\theta \equiv \chi/f_a$. Note that this asymptotic expansion is consistent with the fact that the non-linear evolution disappears in the small-field limit $\theta_\text{e} \rightarrow 0$, namely $\theta_\text{osc} \rightarrow \theta_\text{e}$, in contrast to the approximation used in Ref. \cite{Kawasaki:2011pd}. The asymptotic expansion \eqref{post:asymp} performs better compared to our numerical results of the power spectrum and non-Gaussianity (see the solid blue curve shown in Fig. \ref{fig:fnl}). 

In order to calculate the statistical properties of the final curvature perturbation using the $\delta \mathcal{N}$ formalism \eqref{post:pzeta_dn}, we need to know the total homogeneous e-folding number from $t_\text{e}$ to $t_\text{dec}$, namely $N \equiv \ln{a_\text{dec} \over a_\text{e}}$, and it is divided into two parts,
\be
N = N_\text{osc} + N_\text{dec} ~,
\ee
where $N_\text{osc} = \ln{a_\text{osc} \over a_\text{e}} = {1\over4} \ln{\rho_{r}^\text{e} \over \rho_{r}^\text{osc}}$ and $N_\text{dec} = \ln{a_\text{dec} \over a_\text{osc}} = {1\over4} \ln {\rho_{r}^\text{osc} \over \rho_{r}^\text{dec}}$. Using the Friedmann equation $3 M_\text{Pl}^2 H_\text{osc}^2 = \rho_{r}^\text{osc}$ at the onset of the curvaton oscillation and $\rho_{\chi}^\text{dec} = \rho_{\chi}^\text{osc} e^{-3 N_\text{dec}}$, along with Eqs. \eqref{post:hosc} and \eqref{post:chi_non}, one can calculate the first and second derivatives of the total e-folding number $N$ as
\bl \label{post:deriv1}
N_{,\chi_\text{e}} &= {\alpha \over 3 f_a \theta_\text{osc}} {\sin\theta_\text{osc} \l( 4\theta_\text{osc} + \theta_\text{osc} \cos\theta_\text{osc} + 3 \sin\theta_\text{osc} \r) \over \sin\theta_\text{e} \l( 5 \sin\theta_\text{osc} - \theta_\text{osc} \cos\theta_\text{osc}  \r) } ~,
\\ \label{post:deriv2}
N_{,\chi_\text{e}\chi_\text{e}} &=
{\alpha \over 9 f_a^2}
{ \l(4 \theta_\text{osc} + \theta_\text{osc} \cos\theta_\text{osc} + 3 \sin\theta_\text{osc}  \r)^2 \sin^2\theta_\text{osc} \over \theta_\text{osc}^2 \sin^2\theta_\text{e} (- 5\sin\theta_\text{osc} + \theta_\text{osc} \cos\theta_\text{osc} )^2 }
\nn
\\& \times
\Bigg\{
(1 - \alpha) (\alpha + 3)
+ {1 \over 3 + 4 \theta_\text{osc} \cot{\theta_\text{osc}\over2} - 3 \theta_\text{osc} } \Bigg[
- 3 \theta_{\text{osc}} \csc\theta _{\text{osc}} \cos\theta_\text{e} \left(5-\theta_{\text{osc}} \cot\theta_{\text{osc}} \right) 
\nn \\&
+ 12 \theta_{\text{osc}} \cot\theta_{\text{osc}}
+ 12 \theta_{\text{osc}} \frac{\cot\theta_{\text{osc}} -\theta_{\text{osc}} \csc^2\theta_{\text{osc}}}{5-\theta_{\text{osc}} \cot\theta_{\text{osc}}}
\nn \\&
+ \frac{3 \left(-2 \theta_{\text{osc}}^2-8 \theta_{\text{osc}}^2 \cos\theta_{\text{osc}}+3 \cos(2 \theta_{\text{osc}})-3 \right) \csc\frac{\theta _{\text{osc}}}{2} \sec\frac{\theta_{\text{osc}}}{2}}{ 4 \theta_{\text{osc}} +3 \sin \theta_{\text{osc}} + \theta_{\text{osc}} \cos\theta_{\text{osc}} }
\Bigg]
\Bigg\} ~,
\el
where we defined
\be
\alpha \equiv {3 \rho_\chi \over 3 \rho_\chi + 4 \rho_r } \Big|_\text{dec}
\ee
to characterize the relative density fraction of curvaton at its decay, and manifestly $0 \leq \alpha \leq 1$. It is straightforward to check that if we set the uniform-oscillation condition, namely $\rho_{\chi}^\text{osc} = m_\chi^2 \chi_\text{osc}^2$ and $H_\text{osc} = m_\chi$, the above expressions \eqref{post:deriv1} and \eqref{post:deriv2} reduce to the results found in Ref. \cite{Sasaki:2006kq}: $N_{,\chi_\text{e}} = 2 \alpha g_{,\chi_\text{e}}/(3 g)$ and $N_{,\chi_\text{e}\chi_\text{e}} = 2 \alpha / 9 \l[ \l(3-4\alpha - 2 \alpha^2\r) \l( g_{,\chi_\text{e}}/g \r)^2 + 3 g_{,\chi_\text{e}\chi_\text{e}}/g \r]$.
Note that the $g$ function here can characterize the non-linear evolution of $\chi$ from $t_\text{e}$ to $t_\text{osc}$, which is the case for a non-quadratic potential as we have seen in the left panel of Fig. \ref{fig:nonelov}.

Finally, we derive the final curvature power spectrum $\mathcal{P}_\zeta(t_\text{dec}, k)$ and the non-linear parameter $f_\text{NL}$ at $t_\text{dec}$ using Eqs.~\eqref{post:pzeta_dn} and \eqref{post:fnl_dn}, respectively,
\bl \label{post:pzeta}
\mathcal{P}_\zeta(t_\text{dec}, k) 
&= \mathcal{P}_{\delta\phi}(t_\text{e},k) + ( N_{,\chi_\text{e}} )^2 \mathcal{P}_{\delta\chi}(t_\text{e},k) \nn
\\&=
\mathcal{P}_{\delta\phi}(t_\text{e},k) 
+ \l[ {2 \alpha \over 3 f_a \theta_\text{osc} } F(\theta_\text{e}, \theta_\text{osc}) \r]^2 \mathcal{P}_{\delta\chi}(t_\text{e},k) ~,
\\
f_\text{NL} & = {5\over6} { N_{,\chi\chi} \over \l(N_{,\chi}\r)^2} ~,
\el
where
\be \label{post:ffunc}
F(\theta_\text{e}, \theta_\text{osc}) \equiv {\sin\theta_\text{osc} \l( 4\theta_\text{osc} + \theta_\text{osc} \cos\theta_\text{osc} + 3 \sin\theta_\text{osc} \r) \over 2 \sin\theta_\text{e} \l( 5 \sin\theta_\text{osc} - \theta_\text{osc} \cos\theta_\text{osc} \r) }
\ee
is defined to describe the deviation from the quadratic potential in terms of the non-linear evolution of $\chi$ during the period $(t_\text{e}, t_\text{osc})$ for the axion-like potential \eqref{inf:potential}. This deviation becomes larger when $\chi_e$ approaches the hilltop of the axion-like potential \cite{Kawasaki:2011pd}, as shown by the solid blue curve in the right panel of Fig. \ref{fig:nonelov}. Since the curvaton oscillation epoch will be significantly delayed, causing more contribution to $N_{,\chi_\text{e}}$ and $N_{,\chi_\text{e}\chi_\text{e}}$.
From the left and right panels of Fig. \ref{fig:fnl}, our analytic results (solid blue curves) are comparable to the numerical results (the numerical method will be presented later) and perform better than the analytic formulas used in Ref. \cite{Kobayashi:2020xhm}.
Meanwhile, to satisfy the constraint on $f_\text{NL}$, namely $f_\text{NL} = -0.9 \pm 5.1$ (Planck $68\%$ confidence level) \cite{Planck:2019kim}, $\chi_e$ must be away from $\pi f_a$. It follows from the right panel of Fig. \ref{fig:fnl} that the curvaton field value at the end of inflation lies in the range $0 < \theta_\text{e} \lesssim 3$ for $\alpha \rightarrow 1$. It is also known that $f_\text{NL}$ would be significantly enhanced if the curvaton is still subdominant when it decays \cite{Kawasaki:2011pd}. Hence, the current constraint on $f_\text{NL}$ favors $\alpha \rightarrow 1$, namely the curvaton already dominates when it decays. In addition, the perturbativity condition also suggests the smallness of $f_\text{NL}$ \cite{Kristiano:2021urj, Meng:2022ixx}. 
For the above considerations and simplicity, we do not consider the non-Gaussianity in our model, and the curvaton field value is thus determined from the right panel of Fig. \ref{fig:fnl} such that $\theta_{\text{e}} \simeq 2$. Consequently, there is a slight difference in the peak amplitude of $\mathcal{P}_\zeta(t_\text{dec}, k)$ for the axion-like and quadratic curvaton potentials, see the right panel of Fig. \ref{fig:nonelov} and the left panel of Fig. \ref{fig:fnl}.

It follows from the left panel of Fig. \ref{fig:fnl} and the expression \eqref{post:pzeta}, the peak of the final curvature spectrum $\mathcal{P}_\zeta(t_\text{dec}, k)$ is thus approximately enhanced by a factor $10^{-1} (M_\text{Pl}/f_a)^2$ compared to the peak of $\mathcal{P}_{\delta\chi}(t_\text{e},k)$ shown in Fig. \ref{fig:pchi}. The left panel of Fig. \ref{fig:pchi} also implies that $\mathcal{P}_\zeta(t_\text{dec}, k)$ on large scales will be dominated by $\delta\chi$ if $10^{-1} (M_\text{Pl}/f_a)^2 > 10^3$, namely $f_a < 10^{-2} M_\text{Pl}$, which is favored by the concrete particle models \cite{Kobayashi:2020xhm, Kawasaki:2021ycf}.
Hence, we focus on this parameter region in this paper, namely the final curvature power spectrum on all scales are dominated by the contribution from the curvaton, and consequently, two dips of $\mathcal{P}_{\delta\chi}(t_\text{e},k)$ shown in Fig. \ref{fig:pchi} and Fig. \ref{fig:A} also display to some extent in the final curvature spectra $\mathcal{P}_\zeta(t_\text{dec}, k)$ as shown in Fig. \ref{fig:totpzeta}.

\subsubsection{Numerics of post-inflation dynamics}

\begin{figure}[ht]
	\centering
	\includegraphics[width=0.4\textheight]{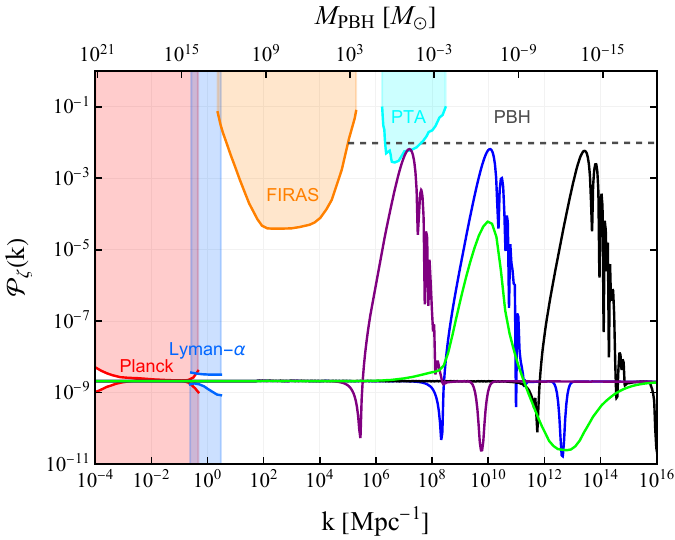}
	\caption{The numerical results of final curvature power spectra $\mathcal{P}_\zeta(t_\text{dec}, k)$ at the curvaton's decay in terms of four set of parameters: $\l\{ \phi_\text{dip}/M_\text{Pl} = 4.8, \sigma_\lambda = 0.01 \r\}$ (the blue curve); $\l\{ \phi_\text{dip}/M_\text{Pl} = 4.5, \sigma_\lambda = 0.01 \r\}$ (the black curve); $\l\{ \phi_\text{dip}/M_\text{Pl} = 5.0, \sigma_\lambda = 0.01 \r\}$ (the purple curve); $\l\{ \phi_\text{dip}/M_\text{Pl} = 4.8, \sigma_\lambda = 0.1 \r\}$ (the green curve), and the rest parameters are the same with Fig. \ref{fig:pchi}.
	The current constraints on $\mathcal{P}_\mathcal{R}(k)$ from Planck \cite{Planck:2018jri}, Lyman-$\alpha$ \cite{Bird:2010mp}, FIRAS \cite{Fixsen:1996nj} and PTA \cite{Byrnes:2018txb} are shown by the shadowed regions, while the grey dashed line refers to $\mathcal{P}_\mathcal{R} \sim 10^{-2}$ in order to produce an abundance of PBHs.}
	\label{fig:totpzeta}
\end{figure}

Employing the sudden-decay approximation, the full background equations are written as
\bl
&3 M_{\mathrm{Pl}}^2 H^2 = \rho_r + \rho_\chi ~, 
\\
&\rho_\chi = {1 \over 2} \dot{\chi}^2 + V_\text{cur}(\chi) ~,
\\
&\dot{\rho}_r + 4 H \rho_r = 0 ~,
\\
&\ddot{\chi} + 3 H \dot{\chi} + V_{\text{cur},\chi} = 0  ~.
\el
Taking the change of variables, namely $N = \ln a(t)$ and $x \equiv m_\chi t$, we obtain two independent equations \cite{Huang:2010cy},
\bl
&N' = \l\{ A e^{-4 N} + B \l[ {1\over2} (\theta')^2 + \tilde{V}(\theta) \r] \r\}^{1/2} ~,
\\
&\theta'' + 3 N' \theta' + {\ddd\tilde{V}(\theta) \over \ddd \theta} = 0 ~,
\el
where the prime denotes the derivative with respect to $x$, and $\tilde{V}(\theta) \equiv V_\text{cur}(\chi)/ (f_a^2 m_\chi^2) = 1 - \cos\theta$, $A \equiv {\rho_{r,e} \over 3 M_{\mathrm{Pl}}^2 m_\chi^2}$ and $B \equiv {f_a^2\over3 M_{\mathrm{Pl}}^2}$. Here $\rho_{r,e}$ is the radiation energy density at the end of inflation. The numerical results of $N_{,\theta_\text{e}}$ and $f_\text{NL}$ are shown by black thick solid curves in Fig. \ref{fig:fnl}. Then, we derive the final curvature spectrum $\mathcal{P}_\zeta(t_\text{dec}, k)$ as shown in Fig. \ref{fig:totpzeta}, in terms of four sets of parameters: $\l\{ \phi_\text{dip}/M_\text{Pl} = 4.8, \sigma_\lambda = 0.01 \r\}$ (the blue curve); $\l\{ \phi_\text{dip}/M_\text{Pl} = 4.5, \sigma_\lambda = 0.01 \r\}$ (the black curve); $\l\{ \phi_\text{dip}/M_\text{Pl} = 5.0, \sigma_\lambda = 0.01 \r\}$ (the purple curve); $\l\{ \phi_\text{dip}/M_\text{Pl} = 4.8, \sigma_\lambda = 0.1 \r\}$ (the green curve), and the rest parameters are the same with Fig. \ref{fig:pchi}.

The spectral index $n_s - 1  \equiv \ddd \ln \mathcal{P}_\zeta / \ddd \ln k$ of $\mathcal{P}_\zeta(t_\text{dec}, k)$ can also derived from the $\delta \mathcal{N}$ formalism \cite{Kawasaki:2011pd, Kobayashi:2020xhm}, 
\be \label{post:ns}
n_s - 1 
\simeq {2 \over 3} {V_{,\chi_\text{e}\chi_\text{e}}(\chi_\text{e}) \over H_\text{inf}^2} + 2 {\dot{H}_\text{inf} \over H_\text{inf}^2}
={2 \over 3} {m_\chi^2 \over H_\text{inf}^2} \cos\theta_\text{e} + 2 {\dot{H}_\text{inf} \over H_\text{inf}^2} ~.
\ee
where we have used the slow-roll approximation during inflation, namely $\ddd\ln k/\ddd t \simeq H_\text{inf}$ at the horizon crossing $k = aH$ and $3 H_\text{inf} \dot{\chi}_\text{e} \simeq - V_{,\chi_\text{e}}(\chi_\text{e})$ at the end of inflation. Since curvaton is light (typically $m_\chi/M_\text{Pl} = 10^{-8}$ in our model), the first term in the right-hand side of Eq. \eqref{post:ns} is thus negligible on CMB scales compared to the second term contributed by Starobinsky inflaton. Hence, our curvaton model is free from the constraint on $n_s$ provided by CMB experiments.

\section{Primordial black holes and scalar-induced gravitational waves} \label{sec:pbh}

\subsection{PBH formation}

The above section shows that the small-scale curvature perturbations can be enhanced in the non-minimal curvaton scenario. 
These large curvature perturbations give rise to overdense regions $\delta \equiv \delta\rho/\rho$ (defined on a comoving uniform-cosmic time slice) during the radiation-dominated era (after the curvaton's decay), leading to gravitationally collapse into PBHs if their sizes are larger than the Jeans scale, or the density perturbation exceeds the threshold $\delta_c$ of the collapse which is taken as $0.51$ in this paper (it is suggested by Ref. \cite{Musco:2020jjb} that $0.4 \lesssim \delta_c \lesssim 0.7$). 
Using the gradient expansion method~\cite{Salopek:1990jq,Deruelle:1994iz,Afshordi:2000nr,Lyth:2004gb} and spherical symmetry, the full non-linear relation between the density perturbation and curvature perturbation on superhorizon scales is given by~\cite{Shibata:1999zs,Kawasaki:2019mbl,Young:2019yug,Kalaja:2019uju,Yoo:2018kvb,DeLuca:2019qsy,Gow:2020bzo,Germani:2018jgr,Musco:2018rwt},
\be \label{pbh:delta}
{\delta\rho \over \rho}
= - {4(1+\omega) \over 5 + 3 \omega} \l({1\over aH}\r)^2 e^{-5 \zeta/2} \nabla^2e^{ \zeta/2}  ~,
\ee
Due to this non-linear relation, even if the curvature perturbation $\zeta$ is Gaussian, the density contrast will not be.
We define the volume-weighted non-linear density contrast (namely the compact function) as~\cite{Musco:2018rwt,Young:2019yug} $\delta_{\rm nl} \equiv {3\over R_m^3} \int_{0}^{R_m} {\delta\rho \over \rho} r^2 \ddd r$, which is related to the linear density perturbation $\delta_l$ as\footnote{The curvature power spectrum $\mathcal{P}_\zeta$ must be larger by a factor of $\mathcal{O}(2)$ to obtain the same PBH abundance, compared with the calculation only based on the linear relation between $\delta$ and $\zeta$~\cite{Young:2019yug}.}
\be
\delta_{\rm nl} = \delta_l - {3\over8} \delta_l^2 ~,
\ee
where $\delta_l$ corresponds to the linear relation ${\delta\rho \over \rho} = - {2(1+\omega) \over 5 + 3 \omega}  \l( {1 \over aH} \r)^2 \nabla^2\zeta$ and $\omega=1/3$ is the parameter of equation of state during the radiation domination. Obviously, $\delta_l$ follows the Gaussian distribution $P(\delta_l) = { 1 \over \sqrt{2 \pi} \sigma_l } \exp\l( -\delta_l^2/(2 \sigma_l^2) \r)$ as $\zeta$ is Gaussian, and the smoothed variance is calculated as $\sigma_l^2 = {1 \over (2\pi)^3} {16\over81} \int \ddd\ln k W(kR)^2 \l({k\over aH}\r)^4 \mathcal{P}_{\zeta}(k)$, where $W(kR)$ is a window function with a smoothing comoving scale $R$ and the mass enclosed by this smoothing volume reads $M \equiv {4\pi\over 3} \rho_c (a R)^3$. In the context of PBH formation, the comoving smoothing scale is customarily chosen as the comoving Hubble radius, namely $R = (a H)^{-1}$. The PBH formation mass is related to the Horizon-crossing $k$ mode of density perturbation, $M \simeq M_\odot \l( k / 1.9 \times 10^6~\text{Mpc}^{-1} \r)^{-2}$ \cite{Sasaki:2018dmp}. In this paper, we choose the real-space top-hat window function:
\be
W(kR) = {1\over (2\pi)^{3/2}} {3 \l[ \sin(kR) - kR \cos(kR) \r] \over (kR)^3} ~.
\ee

\begin{figure}[ht]
	\centering
	\includegraphics[width=0.4\textheight]{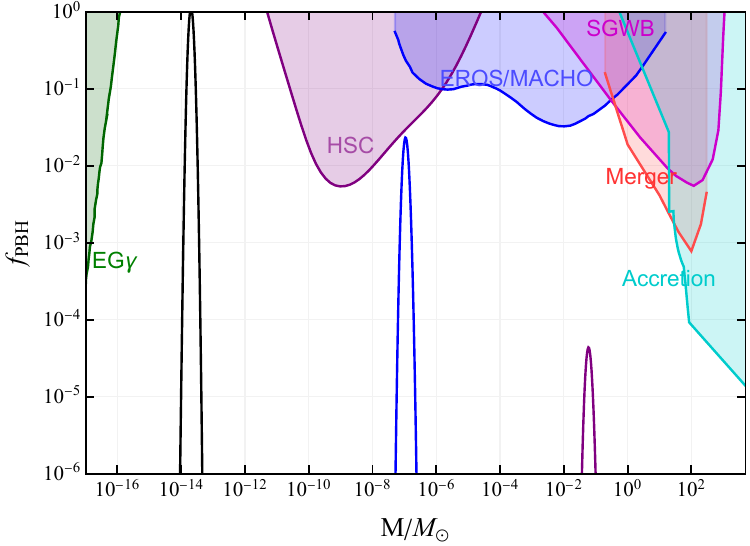}
	\caption{The current PBH abundance $f_\text{PBH}$ in terms of three sets of parameters corresponding to the spectra (the black, blue and purple curves) shown in Fig. \ref{fig:totpzeta}, with various constraints on $f_\text{PBH}$ adopted from Ref. \cite{Carr:2020gox}.}
	\label{fig:fpbh}
\end{figure}

The initial PBH mass function $\beta(M)$ at the formation epoch is defined as $\int \beta(M) \ddd\ln M \equiv \rho_\text{PBH} / \rho_c$, where $\rho_\text{PBH}$ and $\rho_c$ are energy densities of PBHs and radiation background at the PBH formation epoch, respectively. It is customary to estimate $\beta(M)$ using the Press-Schechter formalism \cite{Press:1973iz}, 
\be \label{pbh:beta}
\beta(M) 
= 2 \int_{\delta_c}^{\infty} P(\delta_{\rm nl}) \ddd\delta_{\rm nl} ~.
\ee
Assuming the adiabatic background expansion after PBH formation, one can relate $\beta(M)$ to the current energy fraction \cite{Sasaki:2018dmp} as below,
\be
f_\text{PBH}(M)
\simeq 2.7 \times 10^8 \l( \frac{M}{M_{\odot}} \r)^{-1/2} \beta(M) ~.
\ee
We plot $f_\text{PBH}(M)$ in Fig.~\ref{fig:fpbh} in terms of three spectra (blue, black and purple curves) shown in Fig. \ref{fig:totpzeta}, and the power spectrum shown by the green curve only produce small fraction of PBHs that is not shown in Fig.~\ref{fig:fpbh}. Note that the produced PBH fraction shown by the black curve is able to account for the whole dark matter.

\subsection{Scalar-induced gravitational waves}

Besides the PBHs potentially produced from these enhanced small-scale curvature perturbations, another important physical phenomenon is the generation of sizable GWs, namely the SIGWs. According to the second-order cosmological perturbation theory \cite{Ananda:2006af, Baumann:2007zm}, the second-order tensor modes (i.e., $h_{ij}$ defined in Eq. \eqref{inf:metric_pert}) are inevitably induced through the non-linear coupling between different modes of the first-order scalar modes (i.e., $\Phi$ defined in Eq. \eqref{inf:metric_pert}). SIGWs have been attracting numerous attention in the context of PBH formation (see the comprehensive reviews \cite{Sasaki:2018dmp, Domenech:2021ztg, Escriva:2022duf}), which open a promising GW observational window for detecting PBHs along with the currently operating and forthcoming GW experiments including, for example, SKA \cite{Janssen:2014dka}, Taiji \cite{Ruan:2018tsw},  DECIGO \cite{Kawamura:2011zz}, BBO \cite{Crowder:2005nr},  LISA \cite{Baker:2019nia}, AEDGE \cite{AEDGE:2019nxb}, THEIA \cite{Garcia-Bellido:2021zgu} and $\mu$-ARES \cite{Sesana:2019vho}.

\begin{figure}[ht]
	\centering
	\includegraphics[width=0.4\textheight]{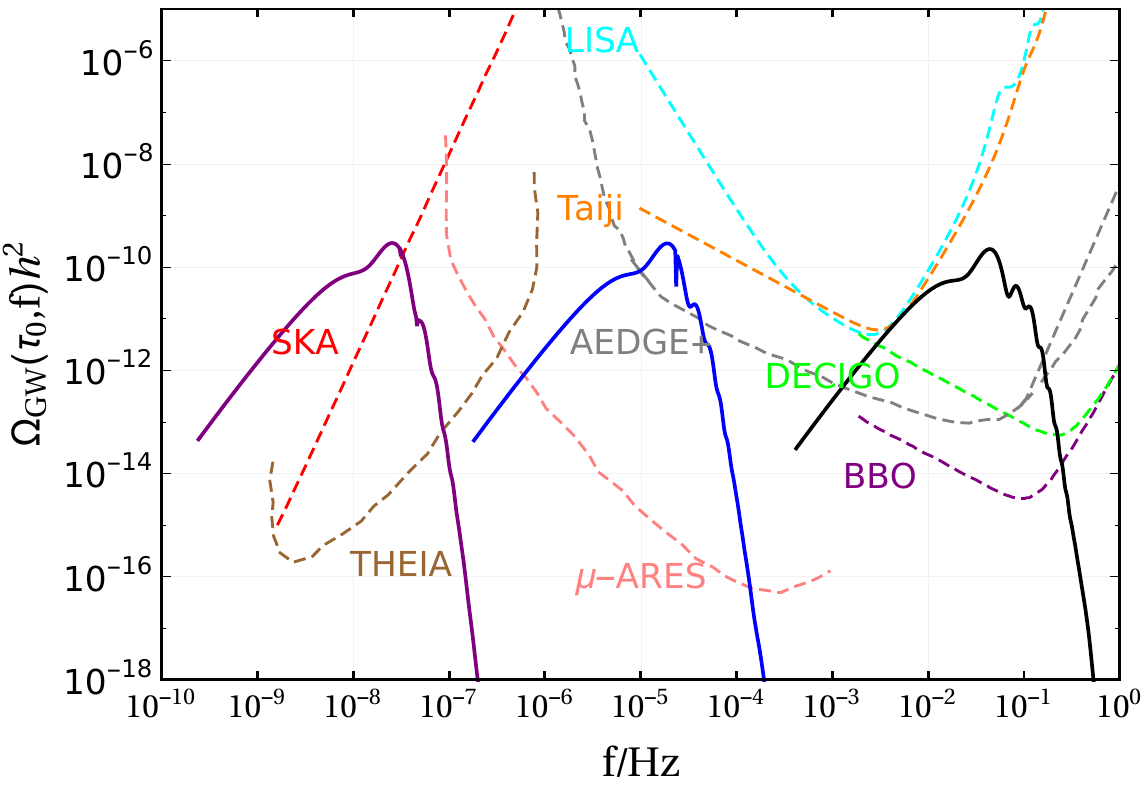}
	\caption{The current energy spectra $\Omega_{\rm GW}(\tau_0, f)$ corresponding to three curvature spectra (the black, blue and purple curves) shown in Fig.~\ref{fig:totpzeta}, along with various GW experiments, e.g. SKA \cite{Janssen:2014dka}, Taiji \cite{Ruan:2018tsw},  DECIGO \cite{Kawamura:2011zz}, BBO \cite{Crowder:2005nr},  LISA \cite{Baker:2019nia}, AEDGE \cite{AEDGE:2019nxb}, THEIA \cite{Garcia-Bellido:2021zgu} and $\mu$-ARES \cite{Sesana:2019vho} .}
	\label{fig:igw}
\end{figure}

The dynamics of SIGWs are given by the second-order perturbative Einstein's field equation,
\be \label{eom_hk}
h_{\textbf{k}}^{\lambda\prime\prime}(\tau) + 2\mathcal{H}h_{\textbf{k}}^{\lambda\prime}(\tau) + k^2h_{\textbf{k}}^{\lambda}(\tau) = S^\lambda_\textbf{k}(\tau) ~,
\ee
where the source term $S^\lambda_\textbf{k}(\tau)$ during the radiation domination is given by
\be
\begin{aligned}
	S^\lambda_{\mathbf{k}}(\tau)
	=
	4 \int {\ddd^3\mathbf{p} \over (2\pi)^{3/2}} \mathbf{e}^\lambda(\mathbf{k},\mathbf{p})
	\Big[&
	3 \Phi_\mathbf{p}(\tau) \Phi_{\mathbf{k} - \mathbf{p}}(\tau)
	+ \mathcal{H}^{-2} \Phi_\mathbf{p}'(\tau) \Phi_{\mathbf{k} - \mathbf{p}}'(\tau)
	\\&
	+ \mathcal{H}^{-1} \Phi_\mathbf{p}'(\tau) \Phi_{\mathbf{k} - \mathbf{p}}(\tau)
	+ \mathcal{H}^{-1} \Phi_\mathbf{p}(\tau) \Phi_{\mathbf{k} - \mathbf{p}}'(\tau)
	\Big] ~,
\end{aligned}
\ee
where $\mathbf{e}^\lambda(\mathbf{k},\mathbf{p}) \equiv e^\lambda_{lm}(\mathbf{k}) p_l p_m$.
Applying the Green function method, namely $h_{\textbf{k}}^{\lambda}(\tau) = \int^\tau \mathrm{d}\tau^\prime g_{k}(\tau,\tau^\prime)S^\lambda_\textbf{k}(\tau^\prime)$, to Eq. \eqref{eom_hk}, one yields the total power spectrum for SIGWs including two polarizations,
\be
\mathcal{P}_h(\tau,k)
= \int^\infty_0 \mathrm{d}v \int^{|1+v|}_{|1-v|}\mathrm{d}u \l[ {4v^2-(1+v^2-u^2)^2 \over 4uv} \r]^2 I^2(v,u,x) \mathcal{P}_\zeta(ku) \mathcal{P}_\zeta(kv) ~.
\ee
where $u \equiv |\mathbf{k} - \mathbf{p}|/k$, $v \equiv p/k$ and $x \equiv k\tau$. The kernel function $I^2(v,u,x)$ describes the source evolution and is calculated at the large-$x$ limit \cite{Kohri:2018awv, Espinosa:2018eve},
\be
\begin{aligned}
	\overline{I^2(v,u,x\rightarrow \infty)} =& 
	4 \left( \frac{3(u^2+v^2-3)}{4 u^3 v^3 x}\right)^2 \bigg[ \bigg(-4uv+(u^2+v^2-3) \ln\left| \frac{3-(u+v)^2}{3-(u-v)^2}\right| \bigg)^2  
	\\&
	+ \pi^2(u^2+v^2-3)^2 \Theta(v+u-\sqrt{3})\bigg] ~,
\end{aligned}
\ee
where the overline denotes the time average. With the canonical definition of GW's energy density \cite{Brill:1964zz, Isaacson:1968hbi, Isaacson:1968zza, Ford:1977dj, Ota:2021fdv}, namely $\rho_{\text{GW}}(\tau,\mathbf{x}) = { M_{\text{pl}}^2 \over 16 a^2(\tau) } \langle h'_{ij}(\tau,\mathbf{x}) h^{ij}{}'(\tau,\mathbf{x}) \rangle$\footnote{Note that the prefactor $1/2$ in the metric perturbations \eqref{inf:metric_pert} is also counted in the total GW energy.}, the GW energy spectrum is written in the following form,
\be \label{omega_gw}
\Omega_{\rm GW}(\tau,k) 
= \frac{1}{48} \l( {k \over a H} \r)^2 \overline{\mathcal{P}_h(\tau,k)} ~.
\ee
The GW energy density starts to decay relative to matter after the radiation-matter equality $\tau_\text{eq}$, the GW spectrum observed today is given by \cite{Pi:2020otn}
\be
\Omega_{\rm GW}(\tau_0, f) h^2 \simeq 1.6 \times 10^{-5} \l( {g_{\ast,s} \over 106.75} \r)^{-1/3} \l( { \Omega_{\rm r,0} h^2 \over 4.1 \times 10^{-5} } \r) \Omega_{\rm GW}(\tau_\text{eq}, f) ~,
\ee
where $\Omega_{\rm GW}(\tau_\text{eq}, f)$ can be calculated by Eq. \eqref{omega_gw} at the radiation-matter equality, with the physical frequency $f = k/(2\pi a_0) \simeq 1.5 \times 10^{-9} (k/1 \text{~pc}^{-1}) ~\text{Hz}$. Since the scalar perturbations damp quickly inside the subhorizon during radiation, a majority of SIGWs is produced just after the source reenters the horizon \cite{Espinosa:2018eve}.  $\Omega_{\rm GW}(\tau_0, f)$ predicted in our model along with various GW sensitivity curves are shown in Fig. \ref{fig:igw}, in terms of three curvature power spectra (the black, blue and purple curves) shown in Fig. \ref{fig:totpzeta}.

\section{Conclusion} \label{sec:con}

In this paper, we revisit the growth of comoving curvature perturbations in non-minimal curvaton scenario. With the detailed analytical and numerical calculations, we identify the role of the non-trivial field metric as an effective friction term (i.e., Eq.~\eqref{inf:etaeff}) for curvaton field perturbations, such that the curvaton field perturbations will grow during inflation when the condition \eqref{inf:etaeff_neg} is satisfied, with the same reason of the enhancement in USR inflation. Besides the superhorizon growth and a dip prior to this growth in the curvaton perturbation spectrum, we also observed a second dip following the enhancement, which arises from the positively large effective friction term. We thus conclude that the growth of curvaton field perturbations during inflation is not merely determined by the depth of the dip but also strongly affected by its first derivative. 
For a case study, we consider a Gaussian-like dip in the field metric. After performing the numerical calculations regarding various depths and widths of the field metric, we confirmed the above conclusion.

Following the formalism presented in Ref.~\cite{Kawasaki:2011pd}, we investigate the post-inflation dynamics in detail and derive the analytical formalism of the curvature power spectrum and non-Gaussianity at the curvaton's decay in terms of the axion-like curvaton potential. Also, we perform the full numerical calculations for both inflationary and post-inflation dynamics to derive the final curvature spectrum and non-Gaussianity. The current PBH abundance in this model is large enough to explain the whole dark matter. The concomitant SIGW signals are detectable by current and forthcoming GW experiments, which serve as a promising approach for detecting PBHs.

\acknowledgments

We thank the anonymous referee for valuable suggestions.
C.C. thanks Chengjie Fu, Sida Lu, Shi Pi, Yu-Cheng Qiu, Misao Sasaki and Tsutomu Yanagida for useful discussion and communication.
C.C. thanks the Particle Cosmology Group at the University of Science and
Technology of China, the Center of Astrophysics at Anhui University, the Center of Astrophysics at Anhui Normal University and Tsung-Dao Lee Institute at Shanghai Jiao Tong University during his visits. 
A.G. thanks Sukannya Bhattacharya and Mayukh Raj Gangopadhyay for the discussion.
Z.L has been supported by the Polish National Science Center grant 2017/27/B/ ST2/02531.
Y.L. is supported by Boya Fellowship of Peking University.
A.N. is supported by ISRO Respond grant. 
This work is supported in part by the National Key R\&D Program of China (Grant No. 2021YFC2203100) and the National SKA Program of China (Grant No. 2020SKA0120100).

\bibliographystyle{apsrev4-1}
\bibliography{alp}

\begin{thebibliography}{98}%
\makeatletter
\providecommand \@ifxundefined [1]{%
 \@ifx{#1\undefined}
}%
\providecommand \@ifnum [1]{%
 \ifnum #1\expandafter \@firstoftwo
 \else \expandafter \@secondoftwo
 \fi
}%
\providecommand \@ifx [1]{%
 \ifx #1\expandafter \@firstoftwo
 \else \expandafter \@secondoftwo
 \fi
}%
\providecommand \natexlab [1]{#1}%
\providecommand \enquote  [1]{``#1''}%
\providecommand \bibnamefont  [1]{#1}%
\providecommand \bibfnamefont [1]{#1}%
\providecommand \citenamefont [1]{#1}%
\providecommand \href@noop [0]{\@secondoftwo}%
\providecommand \href [0]{\begingroup \@sanitize@url \@href}%
\providecommand \@href[1]{\@@startlink{#1}\@@href}%
\providecommand \@@href[1]{\endgroup#1\@@endlink}%
\providecommand \@sanitize@url [0]{\catcode `\\12\catcode `\$12\catcode
  `\&12\catcode `\#12\catcode `\^12\catcode `\_12\catcode `\%12\relax}%
\providecommand \@@startlink[1]{}%
\providecommand \@@endlink[0]{}%
\providecommand \url  [0]{\begingroup\@sanitize@url \@url }%
\providecommand \@url [1]{\endgroup\@href {#1}{\urlprefix }}%
\providecommand \urlprefix  [0]{URL }%
\providecommand \Eprint [0]{\href }%
\providecommand \doibase [0]{http://dx.doi.org/}%
\providecommand \selectlanguage [0]{\@gobble}%
\providecommand \bibinfo  [0]{\@secondoftwo}%
\providecommand \bibfield  [0]{\@secondoftwo}%
\providecommand \translation [1]{[#1]}%
\providecommand \BibitemOpen [0]{}%
\providecommand \bibitemStop [0]{}%
\providecommand \bibitemNoStop [0]{.\EOS\space}%
\providecommand \EOS [0]{\spacefactor3000\relax}%
\providecommand \BibitemShut  [1]{\csname bibitem#1\endcsname}%
\let\auto@bib@innerbib\@empty
\bibitem [{\citenamefont {Akrami}\ \emph
  {et~al.}(2020{\natexlab{a}})\citenamefont {Akrami} \emph
  {et~al.}}]{Planck:2018jri}%
  \BibitemOpen
  \bibfield  {author} {\bibinfo {author} {\bibfnamefont {Y.}~\bibnamefont
  {Akrami}} \emph {et~al.} (\bibinfo {collaboration} {Planck}),\ }\href
  {\doibase 10.1051/0004-6361/201833887} {\bibfield  {journal} {\bibinfo
  {journal} {Astron. Astrophys.}\ }\textbf {\bibinfo {volume} {641}},\ \bibinfo
  {pages} {A10} (\bibinfo {year} {2020}{\natexlab{a}})},\ \Eprint
  {http://arxiv.org/abs/1807.06211} {arXiv:1807.06211 [astro-ph.CO]}
  \BibitemShut {NoStop}%
\bibitem [{\citenamefont {Hawking}(1971)}]{Hawking:1971ei}%
  \BibitemOpen
  \bibfield  {author} {\bibinfo {author} {\bibfnamefont {S.}~\bibnamefont
  {Hawking}},\ }\href@noop {} {\bibfield  {journal} {\bibinfo  {journal} {Mon.
  Not. Roy. Astron. Soc.}\ }\textbf {\bibinfo {volume} {152}},\ \bibinfo
  {pages} {75} (\bibinfo {year} {1971})}\BibitemShut {NoStop}%
\bibitem [{\citenamefont {Carr}\ and\ \citenamefont
  {Hawking}(1974)}]{Carr:1974nx}%
  \BibitemOpen
  \bibfield  {author} {\bibinfo {author} {\bibfnamefont {B.~J.}\ \bibnamefont
  {Carr}}\ and\ \bibinfo {author} {\bibfnamefont {S.~W.}\ \bibnamefont
  {Hawking}},\ }\href@noop {} {\bibfield  {journal} {\bibinfo  {journal} {Mon.
  Not. Roy. Astron. Soc.}\ }\textbf {\bibinfo {volume} {168}},\ \bibinfo
  {pages} {399} (\bibinfo {year} {1974})}\BibitemShut {NoStop}%
\bibitem [{\citenamefont {Carr}(1975)}]{Carr:1975qj}%
  \BibitemOpen
  \bibfield  {author} {\bibinfo {author} {\bibfnamefont {B.~J.}\ \bibnamefont
  {Carr}},\ }\href {\doibase 10.1086/153853} {\bibfield  {journal} {\bibinfo
  {journal} {Astrophys. J.}\ }\textbf {\bibinfo {volume} {201}},\ \bibinfo
  {pages} {1} (\bibinfo {year} {1975})}\BibitemShut {NoStop}%
\bibitem [{\citenamefont {Carr}\ \emph {et~al.}(2010)\citenamefont {Carr},
  \citenamefont {Kohri}, \citenamefont {Sendouda},\ and\ \citenamefont
  {Yokoyama}}]{Carr:2009jm}%
  \BibitemOpen
  \bibfield  {author} {\bibinfo {author} {\bibfnamefont {B.~J.}\ \bibnamefont
  {Carr}}, \bibinfo {author} {\bibfnamefont {K.}~\bibnamefont {Kohri}},
  \bibinfo {author} {\bibfnamefont {Y.}~\bibnamefont {Sendouda}}, \ and\
  \bibinfo {author} {\bibfnamefont {J.}~\bibnamefont {Yokoyama}},\ }\href
  {\doibase 10.1103/PhysRevD.81.104019} {\bibfield  {journal} {\bibinfo
  {journal} {Phys. Rev. D}\ }\textbf {\bibinfo {volume} {81}},\ \bibinfo
  {pages} {104019} (\bibinfo {year} {2010})},\ \Eprint
  {http://arxiv.org/abs/0912.5297} {arXiv:0912.5297 [astro-ph.CO]} \BibitemShut
  {NoStop}%
\bibitem [{\citenamefont {Carr}\ \emph {et~al.}(2021)\citenamefont {Carr},
  \citenamefont {Kohri}, \citenamefont {Sendouda},\ and\ \citenamefont
  {Yokoyama}}]{Carr:2020gox}%
  \BibitemOpen
  \bibfield  {author} {\bibinfo {author} {\bibfnamefont {B.}~\bibnamefont
  {Carr}}, \bibinfo {author} {\bibfnamefont {K.}~\bibnamefont {Kohri}},
  \bibinfo {author} {\bibfnamefont {Y.}~\bibnamefont {Sendouda}}, \ and\
  \bibinfo {author} {\bibfnamefont {J.}~\bibnamefont {Yokoyama}},\ }\href
  {\doibase 10.1088/1361-6633/ac1e31} {\bibfield  {journal} {\bibinfo
  {journal} {Rept. Prog. Phys.}\ }\textbf {\bibinfo {volume} {84}},\ \bibinfo
  {pages} {116902} (\bibinfo {year} {2021})},\ \Eprint
  {http://arxiv.org/abs/2002.12778} {arXiv:2002.12778 [astro-ph.CO]}
  \BibitemShut {NoStop}%
\bibitem [{\citenamefont {Luo}\ \emph {et~al.}(2021)\citenamefont {Luo},
  \citenamefont {Chen}, \citenamefont {Kusakabe},\ and\ \citenamefont
  {Kajino}}]{Luo:2020dlg}%
  \BibitemOpen
  \bibfield  {author} {\bibinfo {author} {\bibfnamefont {Y.}~\bibnamefont
  {Luo}}, \bibinfo {author} {\bibfnamefont {C.}~\bibnamefont {Chen}}, \bibinfo
  {author} {\bibfnamefont {M.}~\bibnamefont {Kusakabe}}, \ and\ \bibinfo
  {author} {\bibfnamefont {T.}~\bibnamefont {Kajino}},\ }\href {\doibase
  10.1088/1475-7516/2021/05/042} {\bibfield  {journal} {\bibinfo  {journal}
  {JCAP}\ }\textbf {\bibinfo {volume} {05}},\ \bibinfo {pages} {042} (\bibinfo
  {year} {2021})},\ \Eprint {http://arxiv.org/abs/2011.10937} {arXiv:2011.10937
  [astro-ph.CO]} \BibitemShut {NoStop}%
\bibitem [{\citenamefont {Cai}\ \emph {et~al.}(2022)\citenamefont {Cai},
  \citenamefont {Chen}, \citenamefont {Ding},\ and\ \citenamefont
  {Wang}}]{Cai:2021zxo}%
  \BibitemOpen
  \bibfield  {author} {\bibinfo {author} {\bibfnamefont {Y.-F.}\ \bibnamefont
  {Cai}}, \bibinfo {author} {\bibfnamefont {C.}~\bibnamefont {Chen}}, \bibinfo
  {author} {\bibfnamefont {Q.}~\bibnamefont {Ding}}, \ and\ \bibinfo {author}
  {\bibfnamefont {Y.}~\bibnamefont {Wang}},\ }\href {\doibase
  10.1140/epjc/s10052-022-10395-w} {\bibfield  {journal} {\bibinfo  {journal}
  {Eur. Phys. J. C}\ }\textbf {\bibinfo {volume} {82}},\ \bibinfo {pages} {464}
  (\bibinfo {year} {2022})},\ \Eprint {http://arxiv.org/abs/2105.11481}
  {arXiv:2105.11481 [astro-ph.CO]} \BibitemShut {NoStop}%
\bibitem [{\citenamefont {Cai}\ \emph {et~al.}(2021)\citenamefont {Cai},
  \citenamefont {Chen}, \citenamefont {Ding},\ and\ \citenamefont
  {Wang}}]{Cai:2021fgm}%
  \BibitemOpen
  \bibfield  {author} {\bibinfo {author} {\bibfnamefont {Y.-F.}\ \bibnamefont
  {Cai}}, \bibinfo {author} {\bibfnamefont {C.}~\bibnamefont {Chen}}, \bibinfo
  {author} {\bibfnamefont {Q.}~\bibnamefont {Ding}}, \ and\ \bibinfo {author}
  {\bibfnamefont {Y.}~\bibnamefont {Wang}},\ }\href@noop {} {\  (\bibinfo
  {year} {2021})},\ \Eprint {http://arxiv.org/abs/2112.10422} {arXiv:2112.10422
  [astro-ph.CO]} \BibitemShut {NoStop}%
\bibitem [{\citenamefont {Bean}\ and\ \citenamefont
  {Magueijo}(2002)}]{Bean:2002kx}%
  \BibitemOpen
  \bibfield  {author} {\bibinfo {author} {\bibfnamefont {R.}~\bibnamefont
  {Bean}}\ and\ \bibinfo {author} {\bibfnamefont {J.}~\bibnamefont
  {Magueijo}},\ }\href {\doibase 10.1103/PhysRevD.66.063505} {\bibfield
  {journal} {\bibinfo  {journal} {Phys. Rev. D}\ }\textbf {\bibinfo {volume}
  {66}},\ \bibinfo {pages} {063505} (\bibinfo {year} {2002})},\ \Eprint
  {http://arxiv.org/abs/astro-ph/0204486} {arXiv:astro-ph/0204486} \BibitemShut
  {NoStop}%
\bibitem [{\citenamefont {Kawasaki}\ \emph {et~al.}(2012)\citenamefont
  {Kawasaki}, \citenamefont {Kusenko},\ and\ \citenamefont
  {Yanagida}}]{Kawasaki:2012kn}%
  \BibitemOpen
  \bibfield  {author} {\bibinfo {author} {\bibfnamefont {M.}~\bibnamefont
  {Kawasaki}}, \bibinfo {author} {\bibfnamefont {A.}~\bibnamefont {Kusenko}}, \
  and\ \bibinfo {author} {\bibfnamefont {T.~T.}\ \bibnamefont {Yanagida}},\
  }\href {\doibase 10.1016/j.physletb.2012.03.056} {\bibfield  {journal}
  {\bibinfo  {journal} {Phys. Lett. B}\ }\textbf {\bibinfo {volume} {711}},\
  \bibinfo {pages} {1} (\bibinfo {year} {2012})},\ \Eprint
  {http://arxiv.org/abs/1202.3848} {arXiv:1202.3848 [astro-ph.CO]} \BibitemShut
  {NoStop}%
\bibitem [{\citenamefont {Yuan}\ \emph {et~al.}(2023)\citenamefont {Yuan},
  \citenamefont {Lei}, \citenamefont {Wang}, \citenamefont {Wang},
  \citenamefont {Wang}, \citenamefont {Chen}, \citenamefont {Shen},
  \citenamefont {Cai},\ and\ \citenamefont {Fan}}]{Yuan:2023bvh}%
  \BibitemOpen
  \bibfield  {author} {\bibinfo {author} {\bibfnamefont {G.-W.}\ \bibnamefont
  {Yuan}}, \bibinfo {author} {\bibfnamefont {L.}~\bibnamefont {Lei}}, \bibinfo
  {author} {\bibfnamefont {Y.-Z.}\ \bibnamefont {Wang}}, \bibinfo {author}
  {\bibfnamefont {B.}~\bibnamefont {Wang}}, \bibinfo {author} {\bibfnamefont
  {Y.-Y.}\ \bibnamefont {Wang}}, \bibinfo {author} {\bibfnamefont
  {C.}~\bibnamefont {Chen}}, \bibinfo {author} {\bibfnamefont {Z.-Q.}\
  \bibnamefont {Shen}}, \bibinfo {author} {\bibfnamefont {Y.-F.}\ \bibnamefont
  {Cai}}, \ and\ \bibinfo {author} {\bibfnamefont {Y.-Z.}\ \bibnamefont
  {Fan}},\ }\href@noop {} {\  (\bibinfo {year} {2023})},\ \Eprint
  {http://arxiv.org/abs/2303.09391} {arXiv:2303.09391 [astro-ph.CO]}
  \BibitemShut {NoStop}%
\bibitem [{\citenamefont {Cai}\ \emph {et~al.}(2023)\citenamefont {Cai},
  \citenamefont {Tang}, \citenamefont {Mo}, \citenamefont {Yan}, \citenamefont
  {Chen}, \citenamefont {Ma}, \citenamefont {Wang}, \citenamefont {Luo},
  \citenamefont {Easson},\ and\ \citenamefont {Marciano}}]{Cai:2023ptf}%
  \BibitemOpen
  \bibfield  {author} {\bibinfo {author} {\bibfnamefont {Y.-F.}\ \bibnamefont
  {Cai}}, \bibinfo {author} {\bibfnamefont {C.}~\bibnamefont {Tang}}, \bibinfo
  {author} {\bibfnamefont {G.}~\bibnamefont {Mo}}, \bibinfo {author}
  {\bibfnamefont {S.}~\bibnamefont {Yan}}, \bibinfo {author} {\bibfnamefont
  {C.}~\bibnamefont {Chen}}, \bibinfo {author} {\bibfnamefont {X.}~\bibnamefont
  {Ma}}, \bibinfo {author} {\bibfnamefont {B.}~\bibnamefont {Wang}}, \bibinfo
  {author} {\bibfnamefont {W.}~\bibnamefont {Luo}}, \bibinfo {author}
  {\bibfnamefont {D.}~\bibnamefont {Easson}}, \ and\ \bibinfo {author}
  {\bibfnamefont {A.}~\bibnamefont {Marciano}},\ }\href@noop {} {\  (\bibinfo
  {year} {2023})},\ \Eprint {http://arxiv.org/abs/2301.09403} {arXiv:2301.09403
  [astro-ph.CO]} \BibitemShut {NoStop}%
\bibitem [{\citenamefont {Bird}\ \emph {et~al.}(2016)\citenamefont {Bird},
  \citenamefont {Cholis}, \citenamefont {Mu\~noz}, \citenamefont
  {Ali-Ha\"\i{}moud}, \citenamefont {Kamionkowski}, \citenamefont {Kovetz},
  \citenamefont {Raccanelli},\ and\ \citenamefont {Riess}}]{Bird:2016dcv}%
  \BibitemOpen
  \bibfield  {author} {\bibinfo {author} {\bibfnamefont {S.}~\bibnamefont
  {Bird}}, \bibinfo {author} {\bibfnamefont {I.}~\bibnamefont {Cholis}},
  \bibinfo {author} {\bibfnamefont {J.~B.}\ \bibnamefont {Mu\~noz}}, \bibinfo
  {author} {\bibfnamefont {Y.}~\bibnamefont {Ali-Ha\"\i{}moud}}, \bibinfo
  {author} {\bibfnamefont {M.}~\bibnamefont {Kamionkowski}}, \bibinfo {author}
  {\bibfnamefont {E.~D.}\ \bibnamefont {Kovetz}}, \bibinfo {author}
  {\bibfnamefont {A.}~\bibnamefont {Raccanelli}}, \ and\ \bibinfo {author}
  {\bibfnamefont {A.~G.}\ \bibnamefont {Riess}},\ }\href {\doibase
  10.1103/PhysRevLett.116.201301} {\bibfield  {journal} {\bibinfo  {journal}
  {Phys. Rev. Lett.}\ }\textbf {\bibinfo {volume} {116}},\ \bibinfo {pages}
  {201301} (\bibinfo {year} {2016})},\ \Eprint
  {http://arxiv.org/abs/1603.00464} {arXiv:1603.00464 [astro-ph.CO]}
  \BibitemShut {NoStop}%
\bibitem [{\citenamefont {Sasaki}\ \emph {et~al.}(2016)\citenamefont {Sasaki},
  \citenamefont {Suyama}, \citenamefont {Tanaka},\ and\ \citenamefont
  {Yokoyama}}]{Sasaki:2016jop}%
  \BibitemOpen
  \bibfield  {author} {\bibinfo {author} {\bibfnamefont {M.}~\bibnamefont
  {Sasaki}}, \bibinfo {author} {\bibfnamefont {T.}~\bibnamefont {Suyama}},
  \bibinfo {author} {\bibfnamefont {T.}~\bibnamefont {Tanaka}}, \ and\ \bibinfo
  {author} {\bibfnamefont {S.}~\bibnamefont {Yokoyama}},\ }\href {\doibase
  10.1103/PhysRevLett.117.061101} {\bibfield  {journal} {\bibinfo  {journal}
  {Phys. Rev. Lett.}\ }\textbf {\bibinfo {volume} {117}},\ \bibinfo {pages}
  {061101} (\bibinfo {year} {2016})},\ \bibinfo {note} {[Erratum:
  Phys.Rev.Lett. 121, 059901 (2018)]},\ \Eprint
  {http://arxiv.org/abs/1603.08338} {arXiv:1603.08338 [astro-ph.CO]}
  \BibitemShut {NoStop}%
\bibitem [{\citenamefont {Jedamzik}(2020)}]{Jedamzik:2020ypm}%
  \BibitemOpen
  \bibfield  {author} {\bibinfo {author} {\bibfnamefont {K.}~\bibnamefont
  {Jedamzik}},\ }\href {\doibase 10.1088/1475-7516/2020/09/022} {\bibfield
  {journal} {\bibinfo  {journal} {JCAP}\ }\textbf {\bibinfo {volume} {09}},\
  \bibinfo {pages} {022} (\bibinfo {year} {2020})},\ \Eprint
  {http://arxiv.org/abs/2006.11172} {arXiv:2006.11172 [astro-ph.CO]}
  \BibitemShut {NoStop}%
\bibitem [{\citenamefont {Carr}\ and\ \citenamefont
  {Kuhnel}(2020)}]{Carr:2020xqk}%
  \BibitemOpen
  \bibfield  {author} {\bibinfo {author} {\bibfnamefont {B.}~\bibnamefont
  {Carr}}\ and\ \bibinfo {author} {\bibfnamefont {F.}~\bibnamefont {Kuhnel}},\
  }\href {\doibase 10.1146/annurev-nucl-050520-125911} {\bibfield  {journal}
  {\bibinfo  {journal} {Ann. Rev. Nucl. Part. Sci.}\ }\textbf {\bibinfo
  {volume} {70}},\ \bibinfo {pages} {355} (\bibinfo {year} {2020})},\ \Eprint
  {http://arxiv.org/abs/2006.02838} {arXiv:2006.02838 [astro-ph.CO]}
  \BibitemShut {NoStop}%
\bibitem [{\citenamefont {Green}\ and\ \citenamefont
  {Kavanagh}(2021)}]{Green:2020jor}%
  \BibitemOpen
  \bibfield  {author} {\bibinfo {author} {\bibfnamefont {A.~M.}\ \bibnamefont
  {Green}}\ and\ \bibinfo {author} {\bibfnamefont {B.~J.}\ \bibnamefont
  {Kavanagh}},\ }\href {\doibase 10.1088/1361-6471/abc534} {\bibfield
  {journal} {\bibinfo  {journal} {J. Phys. G}\ }\textbf {\bibinfo {volume}
  {48}},\ \bibinfo {pages} {043001} (\bibinfo {year} {2021})},\ \Eprint
  {http://arxiv.org/abs/2007.10722} {arXiv:2007.10722 [astro-ph.CO]}
  \BibitemShut {NoStop}%
\bibitem [{\citenamefont {Escriv\`a}\ \emph {et~al.}(2022)\citenamefont
  {Escriv\`a}, \citenamefont {Kuhnel},\ and\ \citenamefont
  {Tada}}]{Escriva:2022duf}%
  \BibitemOpen
  \bibfield  {author} {\bibinfo {author} {\bibfnamefont {A.}~\bibnamefont
  {Escriv\`a}}, \bibinfo {author} {\bibfnamefont {F.}~\bibnamefont {Kuhnel}}, \
  and\ \bibinfo {author} {\bibfnamefont {Y.}~\bibnamefont {Tada}},\ }\href@noop
  {} {\  (\bibinfo {year} {2022})},\ \Eprint {http://arxiv.org/abs/2211.05767}
  {arXiv:2211.05767 [astro-ph.CO]} \BibitemShut {NoStop}%
\bibitem [{\citenamefont {Pi}\ and\ \citenamefont {Sasaki}(2021)}]{Pi:2021dft}%
  \BibitemOpen
  \bibfield  {author} {\bibinfo {author} {\bibfnamefont {S.}~\bibnamefont
  {Pi}}\ and\ \bibinfo {author} {\bibfnamefont {M.}~\bibnamefont {Sasaki}},\
  }\href@noop {} {\  (\bibinfo {year} {2021})},\ \Eprint
  {http://arxiv.org/abs/2112.12680} {arXiv:2112.12680 [astro-ph.CO]}
  \BibitemShut {NoStop}%
\bibitem [{\citenamefont {Meng}\ \emph
  {et~al.}(2022{\natexlab{a}})\citenamefont {Meng}, \citenamefont {Yuan},\ and\
  \citenamefont {Huang}}]{Meng:2022low}%
  \BibitemOpen
  \bibfield  {author} {\bibinfo {author} {\bibfnamefont {D.-S.}\ \bibnamefont
  {Meng}}, \bibinfo {author} {\bibfnamefont {C.}~\bibnamefont {Yuan}}, \ and\
  \bibinfo {author} {\bibfnamefont {Q.-G.}\ \bibnamefont {Huang}},\ }\href@noop
  {} {\  (\bibinfo {year} {2022}{\natexlab{a}})},\ \Eprint
  {http://arxiv.org/abs/2212.03577} {arXiv:2212.03577 [astro-ph.CO]}
  \BibitemShut {NoStop}%
\bibitem [{\citenamefont {Kawasaki}\ \emph {et~al.}(1998)\citenamefont
  {Kawasaki}, \citenamefont {Sugiyama},\ and\ \citenamefont
  {Yanagida}}]{Kawasaki:1997ju}%
  \BibitemOpen
  \bibfield  {author} {\bibinfo {author} {\bibfnamefont {M.}~\bibnamefont
  {Kawasaki}}, \bibinfo {author} {\bibfnamefont {N.}~\bibnamefont {Sugiyama}},
  \ and\ \bibinfo {author} {\bibfnamefont {T.}~\bibnamefont {Yanagida}},\
  }\href {\doibase 10.1103/PhysRevD.57.6050} {\bibfield  {journal} {\bibinfo
  {journal} {Phys. Rev. D}\ }\textbf {\bibinfo {volume} {57}},\ \bibinfo
  {pages} {6050} (\bibinfo {year} {1998})},\ \Eprint
  {http://arxiv.org/abs/hep-ph/9710259} {arXiv:hep-ph/9710259} \BibitemShut
  {NoStop}%
\bibitem [{\citenamefont {Choudhury}\ and\ \citenamefont
  {Mazumdar}(2014)}]{Choudhury:2013woa}%
  \BibitemOpen
  \bibfield  {author} {\bibinfo {author} {\bibfnamefont {S.}~\bibnamefont
  {Choudhury}}\ and\ \bibinfo {author} {\bibfnamefont {A.}~\bibnamefont
  {Mazumdar}},\ }\href {\doibase 10.1016/j.physletb.2014.04.050} {\bibfield
  {journal} {\bibinfo  {journal} {Phys. Lett. B}\ }\textbf {\bibinfo {volume}
  {733}},\ \bibinfo {pages} {270} (\bibinfo {year} {2014})},\ \Eprint
  {http://arxiv.org/abs/1307.5119} {arXiv:1307.5119 [astro-ph.CO]} \BibitemShut
  {NoStop}%
\bibitem [{\citenamefont {Byrnes}\ \emph {et~al.}(2019)\citenamefont {Byrnes},
  \citenamefont {Cole},\ and\ \citenamefont {Patil}}]{Byrnes:2018txb}%
  \BibitemOpen
  \bibfield  {author} {\bibinfo {author} {\bibfnamefont {C.~T.}\ \bibnamefont
  {Byrnes}}, \bibinfo {author} {\bibfnamefont {P.~S.}\ \bibnamefont {Cole}}, \
  and\ \bibinfo {author} {\bibfnamefont {S.~P.}\ \bibnamefont {Patil}},\ }\href
  {\doibase 10.1088/1475-7516/2019/06/028} {\bibfield  {journal} {\bibinfo
  {journal} {JCAP}\ }\textbf {\bibinfo {volume} {06}},\ \bibinfo {pages} {028}
  (\bibinfo {year} {2019})},\ \Eprint {http://arxiv.org/abs/1811.11158}
  {arXiv:1811.11158 [astro-ph.CO]} \BibitemShut {NoStop}%
\bibitem [{\citenamefont {Carrilho}\ \emph {et~al.}(2019)\citenamefont
  {Carrilho}, \citenamefont {Malik},\ and\ \citenamefont
  {Mulryne}}]{Carrilho:2019oqg}%
  \BibitemOpen
  \bibfield  {author} {\bibinfo {author} {\bibfnamefont {P.}~\bibnamefont
  {Carrilho}}, \bibinfo {author} {\bibfnamefont {K.~A.}\ \bibnamefont {Malik}},
  \ and\ \bibinfo {author} {\bibfnamefont {D.~J.}\ \bibnamefont {Mulryne}},\
  }\href {\doibase 10.1103/PhysRevD.100.103529} {\bibfield  {journal} {\bibinfo
   {journal} {Phys. Rev. D}\ }\textbf {\bibinfo {volume} {100}},\ \bibinfo
  {pages} {103529} (\bibinfo {year} {2019})},\ \Eprint
  {http://arxiv.org/abs/1907.05237} {arXiv:1907.05237 [astro-ph.CO]}
  \BibitemShut {NoStop}%
\bibitem [{\citenamefont {Cole}\ \emph {et~al.}(2022)\citenamefont {Cole},
  \citenamefont {Gow}, \citenamefont {Byrnes},\ and\ \citenamefont
  {Patil}}]{Cole:2022xqc}%
  \BibitemOpen
  \bibfield  {author} {\bibinfo {author} {\bibfnamefont {P.~S.}\ \bibnamefont
  {Cole}}, \bibinfo {author} {\bibfnamefont {A.~D.}\ \bibnamefont {Gow}},
  \bibinfo {author} {\bibfnamefont {C.~T.}\ \bibnamefont {Byrnes}}, \ and\
  \bibinfo {author} {\bibfnamefont {S.~P.}\ \bibnamefont {Patil}},\ }\href@noop
  {} {\  (\bibinfo {year} {2022})},\ \Eprint {http://arxiv.org/abs/2204.07573}
  {arXiv:2204.07573 [astro-ph.CO]} \BibitemShut {NoStop}%
\bibitem [{\citenamefont {Dimopoulos}\ \emph {et~al.}(2003)\citenamefont
  {Dimopoulos}, \citenamefont {Lyth}, \citenamefont {Notari},\ and\
  \citenamefont {Riotto}}]{Dimopoulos:2003az}%
  \BibitemOpen
  \bibfield  {author} {\bibinfo {author} {\bibfnamefont {K.}~\bibnamefont
  {Dimopoulos}}, \bibinfo {author} {\bibfnamefont {D.~H.}\ \bibnamefont
  {Lyth}}, \bibinfo {author} {\bibfnamefont {A.}~\bibnamefont {Notari}}, \ and\
  \bibinfo {author} {\bibfnamefont {A.}~\bibnamefont {Riotto}},\ }\href
  {\doibase 10.1088/1126-6708/2003/07/053} {\bibfield  {journal} {\bibinfo
  {journal} {JHEP}\ }\textbf {\bibinfo {volume} {07}},\ \bibinfo {pages} {053}
  (\bibinfo {year} {2003})},\ \Eprint {http://arxiv.org/abs/hep-ph/0304050}
  {arXiv:hep-ph/0304050} \BibitemShut {NoStop}%
\bibitem [{\citenamefont {Kasuya}\ and\ \citenamefont
  {Kawasaki}(2009)}]{Kasuya:2009up}%
  \BibitemOpen
  \bibfield  {author} {\bibinfo {author} {\bibfnamefont {S.}~\bibnamefont
  {Kasuya}}\ and\ \bibinfo {author} {\bibfnamefont {M.}~\bibnamefont
  {Kawasaki}},\ }\href {\doibase 10.1103/PhysRevD.80.023516} {\bibfield
  {journal} {\bibinfo  {journal} {Phys. Rev. D}\ }\textbf {\bibinfo {volume}
  {80}},\ \bibinfo {pages} {023516} (\bibinfo {year} {2009})},\ \Eprint
  {http://arxiv.org/abs/0904.3800} {arXiv:0904.3800 [astro-ph.CO]} \BibitemShut
  {NoStop}%
\bibitem [{\citenamefont {Kawasaki}\ \emph {et~al.}(2013)\citenamefont
  {Kawasaki}, \citenamefont {Kitajima},\ and\ \citenamefont
  {Yanagida}}]{Kawasaki:2012wr}%
  \BibitemOpen
  \bibfield  {author} {\bibinfo {author} {\bibfnamefont {M.}~\bibnamefont
  {Kawasaki}}, \bibinfo {author} {\bibfnamefont {N.}~\bibnamefont {Kitajima}},
  \ and\ \bibinfo {author} {\bibfnamefont {T.~T.}\ \bibnamefont {Yanagida}},\
  }\href {\doibase 10.1103/PhysRevD.87.063519} {\bibfield  {journal} {\bibinfo
  {journal} {Phys. Rev. D}\ }\textbf {\bibinfo {volume} {87}},\ \bibinfo
  {pages} {063519} (\bibinfo {year} {2013})},\ \Eprint
  {http://arxiv.org/abs/1207.2550} {arXiv:1207.2550 [hep-ph]} \BibitemShut
  {NoStop}%
\bibitem [{\citenamefont {Ando}\ \emph
  {et~al.}(2018{\natexlab{a}})\citenamefont {Ando}, \citenamefont {Inomata},
  \citenamefont {Kawasaki}, \citenamefont {Mukaida},\ and\ \citenamefont
  {Yanagida}}]{Ando:2017veq}%
  \BibitemOpen
  \bibfield  {author} {\bibinfo {author} {\bibfnamefont {K.}~\bibnamefont
  {Ando}}, \bibinfo {author} {\bibfnamefont {K.}~\bibnamefont {Inomata}},
  \bibinfo {author} {\bibfnamefont {M.}~\bibnamefont {Kawasaki}}, \bibinfo
  {author} {\bibfnamefont {K.}~\bibnamefont {Mukaida}}, \ and\ \bibinfo
  {author} {\bibfnamefont {T.~T.}\ \bibnamefont {Yanagida}},\ }\href {\doibase
  10.1103/PhysRevD.97.123512} {\bibfield  {journal} {\bibinfo  {journal} {Phys.
  Rev. D}\ }\textbf {\bibinfo {volume} {97}},\ \bibinfo {pages} {123512}
  (\bibinfo {year} {2018}{\natexlab{a}})},\ \Eprint
  {http://arxiv.org/abs/1711.08956} {arXiv:1711.08956 [astro-ph.CO]}
  \BibitemShut {NoStop}%
\bibitem [{\citenamefont {Ando}\ \emph
  {et~al.}(2018{\natexlab{b}})\citenamefont {Ando}, \citenamefont {Kawasaki},\
  and\ \citenamefont {Nakatsuka}}]{Ando:2018nge}%
  \BibitemOpen
  \bibfield  {author} {\bibinfo {author} {\bibfnamefont {K.}~\bibnamefont
  {Ando}}, \bibinfo {author} {\bibfnamefont {M.}~\bibnamefont {Kawasaki}}, \
  and\ \bibinfo {author} {\bibfnamefont {H.}~\bibnamefont {Nakatsuka}},\ }\href
  {\doibase 10.1103/PhysRevD.98.083508} {\bibfield  {journal} {\bibinfo
  {journal} {Phys. Rev. D}\ }\textbf {\bibinfo {volume} {98}},\ \bibinfo
  {pages} {083508} (\bibinfo {year} {2018}{\natexlab{b}})},\ \Eprint
  {http://arxiv.org/abs/1805.07757} {arXiv:1805.07757 [astro-ph.CO]}
  \BibitemShut {NoStop}%
\bibitem [{\citenamefont {Inomata}\ \emph {et~al.}(2021)\citenamefont
  {Inomata}, \citenamefont {Kawasaki}, \citenamefont {Mukaida},\ and\
  \citenamefont {Yanagida}}]{Inomata:2020xad}%
  \BibitemOpen
  \bibfield  {author} {\bibinfo {author} {\bibfnamefont {K.}~\bibnamefont
  {Inomata}}, \bibinfo {author} {\bibfnamefont {M.}~\bibnamefont {Kawasaki}},
  \bibinfo {author} {\bibfnamefont {K.}~\bibnamefont {Mukaida}}, \ and\
  \bibinfo {author} {\bibfnamefont {T.~T.}\ \bibnamefont {Yanagida}},\ }\href
  {\doibase 10.1103/PhysRevLett.126.131301} {\bibfield  {journal} {\bibinfo
  {journal} {Phys. Rev. Lett.}\ }\textbf {\bibinfo {volume} {126}},\ \bibinfo
  {pages} {131301} (\bibinfo {year} {2021})},\ \Eprint
  {http://arxiv.org/abs/2011.01270} {arXiv:2011.01270 [astro-ph.CO]}
  \BibitemShut {NoStop}%
\bibitem [{\citenamefont {Kawasaki}\ and\ \citenamefont
  {Nakatsuka}(2021)}]{Kawasaki:2021ycf}%
  \BibitemOpen
  \bibfield  {author} {\bibinfo {author} {\bibfnamefont {M.}~\bibnamefont
  {Kawasaki}}\ and\ \bibinfo {author} {\bibfnamefont {H.}~\bibnamefont
  {Nakatsuka}},\ }\href {\doibase 10.1088/1475-7516/2021/05/023} {\bibfield
  {journal} {\bibinfo  {journal} {JCAP}\ }\textbf {\bibinfo {volume} {05}},\
  \bibinfo {pages} {023} (\bibinfo {year} {2021})},\ \Eprint
  {http://arxiv.org/abs/2101.11244} {arXiv:2101.11244 [astro-ph.CO]}
  \BibitemShut {NoStop}%
\bibitem [{\citenamefont {Linde}\ and\ \citenamefont
  {Mukhanov}(1997)}]{Linde:1996gt}%
  \BibitemOpen
  \bibfield  {author} {\bibinfo {author} {\bibfnamefont {A.~D.}\ \bibnamefont
  {Linde}}\ and\ \bibinfo {author} {\bibfnamefont {V.~F.}\ \bibnamefont
  {Mukhanov}},\ }\href {\doibase 10.1103/PhysRevD.56.R535} {\bibfield
  {journal} {\bibinfo  {journal} {Phys. Rev. D}\ }\textbf {\bibinfo {volume}
  {56}},\ \bibinfo {pages} {R535} (\bibinfo {year} {1997})},\ \Eprint
  {http://arxiv.org/abs/astro-ph/9610219} {arXiv:astro-ph/9610219} \BibitemShut
  {NoStop}%
\bibitem [{\citenamefont {Lyth}\ and\ \citenamefont
  {Wands}(2002)}]{Lyth:2001nq}%
  \BibitemOpen
  \bibfield  {author} {\bibinfo {author} {\bibfnamefont {D.~H.}\ \bibnamefont
  {Lyth}}\ and\ \bibinfo {author} {\bibfnamefont {D.}~\bibnamefont {Wands}},\
  }\href {\doibase 10.1016/S0370-2693(01)01366-1} {\bibfield  {journal}
  {\bibinfo  {journal} {Phys. Lett. B}\ }\textbf {\bibinfo {volume} {524}},\
  \bibinfo {pages} {5} (\bibinfo {year} {2002})},\ \Eprint
  {http://arxiv.org/abs/hep-ph/0110002} {arXiv:hep-ph/0110002} \BibitemShut
  {NoStop}%
\bibitem [{\citenamefont {Lyth}\ \emph {et~al.}(2003)\citenamefont {Lyth},
  \citenamefont {Ungarelli},\ and\ \citenamefont {Wands}}]{Lyth:2002my}%
  \BibitemOpen
  \bibfield  {author} {\bibinfo {author} {\bibfnamefont {D.~H.}\ \bibnamefont
  {Lyth}}, \bibinfo {author} {\bibfnamefont {C.}~\bibnamefont {Ungarelli}}, \
  and\ \bibinfo {author} {\bibfnamefont {D.}~\bibnamefont {Wands}},\ }\href
  {\doibase 10.1103/PhysRevD.67.023503} {\bibfield  {journal} {\bibinfo
  {journal} {Phys. Rev. D}\ }\textbf {\bibinfo {volume} {67}},\ \bibinfo
  {pages} {023503} (\bibinfo {year} {2003})},\ \Eprint
  {http://arxiv.org/abs/astro-ph/0208055} {arXiv:astro-ph/0208055} \BibitemShut
  {NoStop}%
\bibitem [{\citenamefont {Malik}\ \emph {et~al.}(2003)\citenamefont {Malik},
  \citenamefont {Wands},\ and\ \citenamefont {Ungarelli}}]{Malik:2002jb}%
  \BibitemOpen
  \bibfield  {author} {\bibinfo {author} {\bibfnamefont {K.~A.}\ \bibnamefont
  {Malik}}, \bibinfo {author} {\bibfnamefont {D.}~\bibnamefont {Wands}}, \ and\
  \bibinfo {author} {\bibfnamefont {C.}~\bibnamefont {Ungarelli}},\ }\href
  {\doibase 10.1103/PhysRevD.67.063516} {\bibfield  {journal} {\bibinfo
  {journal} {Phys. Rev. D}\ }\textbf {\bibinfo {volume} {67}},\ \bibinfo
  {pages} {063516} (\bibinfo {year} {2003})},\ \Eprint
  {http://arxiv.org/abs/astro-ph/0211602} {arXiv:astro-ph/0211602} \BibitemShut
  {NoStop}%
\bibitem [{\citenamefont {Malik}\ and\ \citenamefont
  {Lyth}(2006)}]{Malik:2006pm}%
  \BibitemOpen
  \bibfield  {author} {\bibinfo {author} {\bibfnamefont {K.~A.}\ \bibnamefont
  {Malik}}\ and\ \bibinfo {author} {\bibfnamefont {D.~H.}\ \bibnamefont
  {Lyth}},\ }\href {\doibase 10.1088/1475-7516/2006/09/008} {\bibfield
  {journal} {\bibinfo  {journal} {JCAP}\ }\textbf {\bibinfo {volume} {09}},\
  \bibinfo {pages} {008} (\bibinfo {year} {2006})},\ \Eprint
  {http://arxiv.org/abs/astro-ph/0604387} {arXiv:astro-ph/0604387} \BibitemShut
  {NoStop}%
\bibitem [{\citenamefont {Kawasaki}\ \emph {et~al.}(2011)\citenamefont
  {Kawasaki}, \citenamefont {Kobayashi},\ and\ \citenamefont
  {Takahashi}}]{Kawasaki:2011pd}%
  \BibitemOpen
  \bibfield  {author} {\bibinfo {author} {\bibfnamefont {M.}~\bibnamefont
  {Kawasaki}}, \bibinfo {author} {\bibfnamefont {T.}~\bibnamefont {Kobayashi}},
  \ and\ \bibinfo {author} {\bibfnamefont {F.}~\bibnamefont {Takahashi}},\
  }\href {\doibase 10.1103/PhysRevD.84.123506} {\bibfield  {journal} {\bibinfo
  {journal} {Phys. Rev. D}\ }\textbf {\bibinfo {volume} {84}},\ \bibinfo
  {pages} {123506} (\bibinfo {year} {2011})},\ \Eprint
  {http://arxiv.org/abs/1107.6011} {arXiv:1107.6011 [astro-ph.CO]} \BibitemShut
  {NoStop}%
\bibitem [{\citenamefont {Gomez-Reino}\ and\ \citenamefont
  {Scrucca}(2006)}]{Gomez-Reino:2006sqc}%
  \BibitemOpen
  \bibfield  {author} {\bibinfo {author} {\bibfnamefont {M.}~\bibnamefont
  {Gomez-Reino}}\ and\ \bibinfo {author} {\bibfnamefont {C.~A.}\ \bibnamefont
  {Scrucca}},\ }\href {\doibase 10.1088/1126-6708/2006/09/008} {\bibfield
  {journal} {\bibinfo  {journal} {JHEP}\ }\textbf {\bibinfo {volume} {09}},\
  \bibinfo {pages} {008} (\bibinfo {year} {2006})},\ \Eprint
  {http://arxiv.org/abs/hep-th/0606273} {arXiv:hep-th/0606273} \BibitemShut
  {NoStop}%
\bibitem [{\citenamefont {Covi}\ \emph
  {et~al.}(2008{\natexlab{a}})\citenamefont {Covi}, \citenamefont
  {Gomez-Reino}, \citenamefont {Gross}, \citenamefont {Louis}, \citenamefont
  {Palma},\ and\ \citenamefont {Scrucca}}]{Covi:2008ea}%
  \BibitemOpen
  \bibfield  {author} {\bibinfo {author} {\bibfnamefont {L.}~\bibnamefont
  {Covi}}, \bibinfo {author} {\bibfnamefont {M.}~\bibnamefont {Gomez-Reino}},
  \bibinfo {author} {\bibfnamefont {C.}~\bibnamefont {Gross}}, \bibinfo
  {author} {\bibfnamefont {J.}~\bibnamefont {Louis}}, \bibinfo {author}
  {\bibfnamefont {G.~A.}\ \bibnamefont {Palma}}, \ and\ \bibinfo {author}
  {\bibfnamefont {C.~A.}\ \bibnamefont {Scrucca}},\ }\href {\doibase
  10.1088/1126-6708/2008/06/057} {\bibfield  {journal} {\bibinfo  {journal}
  {JHEP}\ }\textbf {\bibinfo {volume} {06}},\ \bibinfo {pages} {057} (\bibinfo
  {year} {2008}{\natexlab{a}})},\ \Eprint {http://arxiv.org/abs/0804.1073}
  {arXiv:0804.1073 [hep-th]} \BibitemShut {NoStop}%
\bibitem [{\citenamefont {Covi}\ \emph
  {et~al.}(2008{\natexlab{b}})\citenamefont {Covi}, \citenamefont
  {Gomez-Reino}, \citenamefont {Gross}, \citenamefont {Louis}, \citenamefont
  {Palma},\ and\ \citenamefont {Scrucca}}]{Covi:2008cn}%
  \BibitemOpen
  \bibfield  {author} {\bibinfo {author} {\bibfnamefont {L.}~\bibnamefont
  {Covi}}, \bibinfo {author} {\bibfnamefont {M.}~\bibnamefont {Gomez-Reino}},
  \bibinfo {author} {\bibfnamefont {C.}~\bibnamefont {Gross}}, \bibinfo
  {author} {\bibfnamefont {J.}~\bibnamefont {Louis}}, \bibinfo {author}
  {\bibfnamefont {G.~A.}\ \bibnamefont {Palma}}, \ and\ \bibinfo {author}
  {\bibfnamefont {C.~A.}\ \bibnamefont {Scrucca}},\ }\href {\doibase
  10.1088/1126-6708/2008/08/055} {\bibfield  {journal} {\bibinfo  {journal}
  {JHEP}\ }\textbf {\bibinfo {volume} {08}},\ \bibinfo {pages} {055} (\bibinfo
  {year} {2008}{\natexlab{b}})},\ \Eprint {http://arxiv.org/abs/0805.3290}
  {arXiv:0805.3290 [hep-th]} \BibitemShut {NoStop}%
\bibitem [{\citenamefont {Lalak}\ \emph {et~al.}(2007)\citenamefont {Lalak},
  \citenamefont {Langlois}, \citenamefont {Pokorski},\ and\ \citenamefont
  {Turzynski}}]{Lalak:2007vi}%
  \BibitemOpen
  \bibfield  {author} {\bibinfo {author} {\bibfnamefont {Z.}~\bibnamefont
  {Lalak}}, \bibinfo {author} {\bibfnamefont {D.}~\bibnamefont {Langlois}},
  \bibinfo {author} {\bibfnamefont {S.}~\bibnamefont {Pokorski}}, \ and\
  \bibinfo {author} {\bibfnamefont {K.}~\bibnamefont {Turzynski}},\ }\href
  {\doibase 10.1088/1475-7516/2007/07/014} {\bibfield  {journal} {\bibinfo
  {journal} {JCAP}\ }\textbf {\bibinfo {volume} {07}},\ \bibinfo {pages} {014}
  (\bibinfo {year} {2007})},\ \Eprint {http://arxiv.org/abs/0704.0212}
  {arXiv:0704.0212 [hep-th]} \BibitemShut {NoStop}%
\bibitem [{\citenamefont {Armendariz-Picon}\ \emph {et~al.}(1999)\citenamefont
  {Armendariz-Picon}, \citenamefont {Damour},\ and\ \citenamefont
  {Mukhanov}}]{Armendariz-Picon:1999hyi}%
  \BibitemOpen
  \bibfield  {author} {\bibinfo {author} {\bibfnamefont {C.}~\bibnamefont
  {Armendariz-Picon}}, \bibinfo {author} {\bibfnamefont {T.}~\bibnamefont
  {Damour}}, \ and\ \bibinfo {author} {\bibfnamefont {V.~F.}\ \bibnamefont
  {Mukhanov}},\ }\href {\doibase 10.1016/S0370-2693(99)00603-6} {\bibfield
  {journal} {\bibinfo  {journal} {Phys. Lett. B}\ }\textbf {\bibinfo {volume}
  {458}},\ \bibinfo {pages} {209} (\bibinfo {year} {1999})},\ \Eprint
  {http://arxiv.org/abs/hep-th/9904075} {arXiv:hep-th/9904075} \BibitemShut
  {NoStop}%
\bibitem [{\citenamefont {Armendariz-Picon}\ \emph {et~al.}(2001)\citenamefont
  {Armendariz-Picon}, \citenamefont {Mukhanov},\ and\ \citenamefont
  {Steinhardt}}]{Armendariz-Picon:2000ulo}%
  \BibitemOpen
  \bibfield  {author} {\bibinfo {author} {\bibfnamefont {C.}~\bibnamefont
  {Armendariz-Picon}}, \bibinfo {author} {\bibfnamefont {V.~F.}\ \bibnamefont
  {Mukhanov}}, \ and\ \bibinfo {author} {\bibfnamefont {P.~J.}\ \bibnamefont
  {Steinhardt}},\ }\href {\doibase 10.1103/PhysRevD.63.103510} {\bibfield
  {journal} {\bibinfo  {journal} {Phys. Rev. D}\ }\textbf {\bibinfo {volume}
  {63}},\ \bibinfo {pages} {103510} (\bibinfo {year} {2001})},\ \Eprint
  {http://arxiv.org/abs/astro-ph/0006373} {arXiv:astro-ph/0006373} \BibitemShut
  {NoStop}%
\bibitem [{\citenamefont {Dom\`enech}\ and\ \citenamefont
  {Sasaki}(2015)}]{Domenech:2015qoa}%
  \BibitemOpen
  \bibfield  {author} {\bibinfo {author} {\bibfnamefont {G.}~\bibnamefont
  {Dom\`enech}}\ and\ \bibinfo {author} {\bibfnamefont {M.}~\bibnamefont
  {Sasaki}},\ }\href {\doibase 10.1088/1475-7516/2015/04/022} {\bibfield
  {journal} {\bibinfo  {journal} {JCAP}\ }\textbf {\bibinfo {volume} {04}},\
  \bibinfo {pages} {022} (\bibinfo {year} {2015})},\ \Eprint
  {http://arxiv.org/abs/1501.07699} {arXiv:1501.07699 [gr-qc]} \BibitemShut
  {NoStop}%
\bibitem [{\citenamefont {De~Angelis}\ and\ \citenamefont {van~de
  Bruck}(2023)}]{DeAngelis:2023fdu}%
  \BibitemOpen
  \bibfield  {author} {\bibinfo {author} {\bibfnamefont {M.}~\bibnamefont
  {De~Angelis}}\ and\ \bibinfo {author} {\bibfnamefont {C.}~\bibnamefont
  {van~de Bruck}},\ }\href@noop {} {\  (\bibinfo {year} {2023})},\ \Eprint
  {http://arxiv.org/abs/2304.12364} {arXiv:2304.12364 [hep-th]} \BibitemShut
  {NoStop}%
\bibitem [{\citenamefont {Lola}\ \emph {et~al.}(2021)\citenamefont {Lola},
  \citenamefont {Lymperis},\ and\ \citenamefont {Saridakis}}]{Lola:2020lvk}%
  \BibitemOpen
  \bibfield  {author} {\bibinfo {author} {\bibfnamefont {S.}~\bibnamefont
  {Lola}}, \bibinfo {author} {\bibfnamefont {A.}~\bibnamefont {Lymperis}}, \
  and\ \bibinfo {author} {\bibfnamefont {E.~N.}\ \bibnamefont {Saridakis}},\
  }\href {\doibase 10.1140/epjc/s10052-021-09516-8} {\bibfield  {journal}
  {\bibinfo  {journal} {Eur. Phys. J. C}\ }\textbf {\bibinfo {volume} {81}},\
  \bibinfo {pages} {719} (\bibinfo {year} {2021})},\ \Eprint
  {http://arxiv.org/abs/2005.14069} {arXiv:2005.14069 [gr-qc]} \BibitemShut
  {NoStop}%
\bibitem [{\citenamefont {Choudhury}\ \emph
  {et~al.}(2023{\natexlab{a}})\citenamefont {Choudhury}, \citenamefont
  {Panda},\ and\ \citenamefont {Sami}}]{Choudhury:2023jlt}%
  \BibitemOpen
  \bibfield  {author} {\bibinfo {author} {\bibfnamefont {S.}~\bibnamefont
  {Choudhury}}, \bibinfo {author} {\bibfnamefont {S.}~\bibnamefont {Panda}}, \
  and\ \bibinfo {author} {\bibfnamefont {M.}~\bibnamefont {Sami}},\ }\href@noop
  {} {\  (\bibinfo {year} {2023}{\natexlab{a}})},\ \Eprint
  {http://arxiv.org/abs/2302.05655} {arXiv:2302.05655 [astro-ph.CO]}
  \BibitemShut {NoStop}%
\bibitem [{\citenamefont {Choudhury}\ \emph
  {et~al.}(2023{\natexlab{b}})\citenamefont {Choudhury}, \citenamefont
  {Panda},\ and\ \citenamefont {Sami}}]{Choudhury:2023hvf}%
  \BibitemOpen
  \bibfield  {author} {\bibinfo {author} {\bibfnamefont {S.}~\bibnamefont
  {Choudhury}}, \bibinfo {author} {\bibfnamefont {S.}~\bibnamefont {Panda}}, \
  and\ \bibinfo {author} {\bibfnamefont {M.}~\bibnamefont {Sami}},\ }\href@noop
  {} {\  (\bibinfo {year} {2023}{\natexlab{b}})},\ \Eprint
  {http://arxiv.org/abs/2304.04065} {arXiv:2304.04065 [astro-ph.CO]}
  \BibitemShut {NoStop}%
\bibitem [{\citenamefont {Starobinsky}(1985)}]{Starobinsky:1985ibc}%
  \BibitemOpen
  \bibfield  {author} {\bibinfo {author} {\bibfnamefont {A.~A.}\ \bibnamefont
  {Starobinsky}},\ }\href@noop {} {\bibfield  {journal} {\bibinfo  {journal}
  {JETP Lett.}\ }\textbf {\bibinfo {volume} {42}},\ \bibinfo {pages} {152}
  (\bibinfo {year} {1985})}\BibitemShut {NoStop}%
\bibitem [{\citenamefont {Sasaki}\ and\ \citenamefont
  {Stewart}(1996)}]{Sasaki:1995aw}%
  \BibitemOpen
  \bibfield  {author} {\bibinfo {author} {\bibfnamefont {M.}~\bibnamefont
  {Sasaki}}\ and\ \bibinfo {author} {\bibfnamefont {E.~D.}\ \bibnamefont
  {Stewart}},\ }\href {\doibase 10.1143/PTP.95.71} {\bibfield  {journal}
  {\bibinfo  {journal} {Prog. Theor. Phys.}\ }\textbf {\bibinfo {volume}
  {95}},\ \bibinfo {pages} {71} (\bibinfo {year} {1996})},\ \Eprint
  {http://arxiv.org/abs/astro-ph/9507001} {arXiv:astro-ph/9507001} \BibitemShut
  {NoStop}%
\bibitem [{\citenamefont {Lyth}\ \emph {et~al.}(2005)\citenamefont {Lyth},
  \citenamefont {Malik},\ and\ \citenamefont {Sasaki}}]{Lyth:2004gb}%
  \BibitemOpen
  \bibfield  {author} {\bibinfo {author} {\bibfnamefont {D.~H.}\ \bibnamefont
  {Lyth}}, \bibinfo {author} {\bibfnamefont {K.~A.}\ \bibnamefont {Malik}}, \
  and\ \bibinfo {author} {\bibfnamefont {M.}~\bibnamefont {Sasaki}},\ }\href
  {\doibase 10.1088/1475-7516/2005/05/004} {\bibfield  {journal} {\bibinfo
  {journal} {JCAP}\ }\textbf {\bibinfo {volume} {05}},\ \bibinfo {pages} {004}
  (\bibinfo {year} {2005})},\ \Eprint {http://arxiv.org/abs/astro-ph/0411220}
  {arXiv:astro-ph/0411220} \BibitemShut {NoStop}%
\bibitem [{\citenamefont {Abolhasani}\ \emph {et~al.}(2019)\citenamefont
  {Abolhasani}, \citenamefont {Firouzjahi}, \citenamefont {Naruko},\ and\
  \citenamefont {Sasaki}}]{Abolhasani:2019cqw}%
  \BibitemOpen
  \bibfield  {author} {\bibinfo {author} {\bibfnamefont {A.~A.}\ \bibnamefont
  {Abolhasani}}, \bibinfo {author} {\bibfnamefont {H.}~\bibnamefont
  {Firouzjahi}}, \bibinfo {author} {\bibfnamefont {A.}~\bibnamefont {Naruko}},
  \ and\ \bibinfo {author} {\bibfnamefont {M.}~\bibnamefont {Sasaki}},\ }\href
  {\doibase 10.1142/10953} {\emph {\bibinfo {title} {{Delta N Formalism in
  Cosmological Perturbation Theory}}}}\ (\bibinfo  {publisher} {WSP},\ \bibinfo
  {year} {2019})\BibitemShut {NoStop}%
\bibitem [{\citenamefont {Sasaki}\ \emph {et~al.}(2006)\citenamefont {Sasaki},
  \citenamefont {Valiviita},\ and\ \citenamefont {Wands}}]{Sasaki:2006kq}%
  \BibitemOpen
  \bibfield  {author} {\bibinfo {author} {\bibfnamefont {M.}~\bibnamefont
  {Sasaki}}, \bibinfo {author} {\bibfnamefont {J.}~\bibnamefont {Valiviita}}, \
  and\ \bibinfo {author} {\bibfnamefont {D.}~\bibnamefont {Wands}},\ }\href
  {\doibase 10.1103/PhysRevD.74.103003} {\bibfield  {journal} {\bibinfo
  {journal} {Phys. Rev. D}\ }\textbf {\bibinfo {volume} {74}},\ \bibinfo
  {pages} {103003} (\bibinfo {year} {2006})},\ \Eprint
  {http://arxiv.org/abs/astro-ph/0607627} {arXiv:astro-ph/0607627} \BibitemShut
  {NoStop}%
\bibitem [{\citenamefont {Ichikawa}\ \emph {et~al.}(2008)\citenamefont
  {Ichikawa}, \citenamefont {Suyama}, \citenamefont {Takahashi},\ and\
  \citenamefont {Yamaguchi}}]{Ichikawa:2008iq}%
  \BibitemOpen
  \bibfield  {author} {\bibinfo {author} {\bibfnamefont {K.}~\bibnamefont
  {Ichikawa}}, \bibinfo {author} {\bibfnamefont {T.}~\bibnamefont {Suyama}},
  \bibinfo {author} {\bibfnamefont {T.}~\bibnamefont {Takahashi}}, \ and\
  \bibinfo {author} {\bibfnamefont {M.}~\bibnamefont {Yamaguchi}},\ }\href
  {\doibase 10.1103/PhysRevD.78.023513} {\bibfield  {journal} {\bibinfo
  {journal} {Phys. Rev. D}\ }\textbf {\bibinfo {volume} {78}},\ \bibinfo
  {pages} {023513} (\bibinfo {year} {2008})},\ \Eprint
  {http://arxiv.org/abs/0802.4138} {arXiv:0802.4138 [astro-ph]} \BibitemShut
  {NoStop}%
\bibitem [{\citenamefont {Kobayashi}(2020)}]{Kobayashi:2020xhm}%
  \BibitemOpen
  \bibfield  {author} {\bibinfo {author} {\bibfnamefont {T.}~\bibnamefont
  {Kobayashi}},\ }\href {\doibase 10.1103/PhysRevLett.125.011302} {\bibfield
  {journal} {\bibinfo  {journal} {Phys. Rev. Lett.}\ }\textbf {\bibinfo
  {volume} {125}},\ \bibinfo {pages} {011302} (\bibinfo {year} {2020})},\
  \Eprint {http://arxiv.org/abs/2005.01741} {arXiv:2005.01741 [astro-ph.CO]}
  \BibitemShut {NoStop}%
\bibitem [{\citenamefont {Maldacena}(2003)}]{Maldacena:2002vr}%
  \BibitemOpen
  \bibfield  {author} {\bibinfo {author} {\bibfnamefont {J.~M.}\ \bibnamefont
  {Maldacena}},\ }\href {\doibase 10.1088/1126-6708/2003/05/013} {\bibfield
  {journal} {\bibinfo  {journal} {JHEP}\ }\textbf {\bibinfo {volume} {05}},\
  \bibinfo {pages} {013} (\bibinfo {year} {2003})},\ \Eprint
  {http://arxiv.org/abs/astro-ph/0210603} {arXiv:astro-ph/0210603} \BibitemShut
  {NoStop}%
\bibitem [{\citenamefont {Akrami}\ \emph
  {et~al.}(2020{\natexlab{b}})\citenamefont {Akrami} \emph
  {et~al.}}]{Planck:2019kim}%
  \BibitemOpen
  \bibfield  {author} {\bibinfo {author} {\bibfnamefont {Y.}~\bibnamefont
  {Akrami}} \emph {et~al.} (\bibinfo {collaboration} {Planck}),\ }\href
  {\doibase 10.1051/0004-6361/201935891} {\bibfield  {journal} {\bibinfo
  {journal} {Astron. Astrophys.}\ }\textbf {\bibinfo {volume} {641}},\ \bibinfo
  {pages} {A9} (\bibinfo {year} {2020}{\natexlab{b}})},\ \Eprint
  {http://arxiv.org/abs/1905.05697} {arXiv:1905.05697 [astro-ph.CO]}
  \BibitemShut {NoStop}%
\bibitem [{\citenamefont {Kristiano}\ and\ \citenamefont
  {Yokoyama}(2022)}]{Kristiano:2021urj}%
  \BibitemOpen
  \bibfield  {author} {\bibinfo {author} {\bibfnamefont {J.}~\bibnamefont
  {Kristiano}}\ and\ \bibinfo {author} {\bibfnamefont {J.}~\bibnamefont
  {Yokoyama}},\ }\href {\doibase 10.1103/PhysRevLett.128.061301} {\bibfield
  {journal} {\bibinfo  {journal} {Phys. Rev. Lett.}\ }\textbf {\bibinfo
  {volume} {128}},\ \bibinfo {pages} {061301} (\bibinfo {year} {2022})},\
  \Eprint {http://arxiv.org/abs/2104.01953} {arXiv:2104.01953 [hep-th]}
  \BibitemShut {NoStop}%
\bibitem [{\citenamefont {Meng}\ \emph
  {et~al.}(2022{\natexlab{b}})\citenamefont {Meng}, \citenamefont {Yuan},\ and\
  \citenamefont {Huang}}]{Meng:2022ixx}%
  \BibitemOpen
  \bibfield  {author} {\bibinfo {author} {\bibfnamefont {D.-S.}\ \bibnamefont
  {Meng}}, \bibinfo {author} {\bibfnamefont {C.}~\bibnamefont {Yuan}}, \ and\
  \bibinfo {author} {\bibfnamefont {Q.-g.}\ \bibnamefont {Huang}},\ }\href
  {\doibase 10.1103/PhysRevD.106.063508} {\bibfield  {journal} {\bibinfo
  {journal} {Phys. Rev. D}\ }\textbf {\bibinfo {volume} {106}},\ \bibinfo
  {pages} {063508} (\bibinfo {year} {2022}{\natexlab{b}})},\ \Eprint
  {http://arxiv.org/abs/2207.07668} {arXiv:2207.07668 [astro-ph.CO]}
  \BibitemShut {NoStop}%
\bibitem [{\citenamefont {Huang}(2010)}]{Huang:2010cy}%
  \BibitemOpen
  \bibfield  {author} {\bibinfo {author} {\bibfnamefont {Q.-G.}\ \bibnamefont
  {Huang}},\ }\href {\doibase 10.1088/1475-7516/2011/02/E01} {\bibfield
  {journal} {\bibinfo  {journal} {JCAP}\ }\textbf {\bibinfo {volume} {11}},\
  \bibinfo {pages} {026} (\bibinfo {year} {2010})},\ \bibinfo {note} {[Erratum:
  JCAP 02, E01 (2011)]},\ \Eprint {http://arxiv.org/abs/1008.2641}
  {arXiv:1008.2641 [astro-ph.CO]} \BibitemShut {NoStop}%
\bibitem [{\citenamefont {Bird}\ \emph {et~al.}(2011)\citenamefont {Bird},
  \citenamefont {Peiris}, \citenamefont {Viel},\ and\ \citenamefont
  {Verde}}]{Bird:2010mp}%
  \BibitemOpen
  \bibfield  {author} {\bibinfo {author} {\bibfnamefont {S.}~\bibnamefont
  {Bird}}, \bibinfo {author} {\bibfnamefont {H.~V.}\ \bibnamefont {Peiris}},
  \bibinfo {author} {\bibfnamefont {M.}~\bibnamefont {Viel}}, \ and\ \bibinfo
  {author} {\bibfnamefont {L.}~\bibnamefont {Verde}},\ }\href {\doibase
  10.1111/j.1365-2966.2011.18245.x} {\bibfield  {journal} {\bibinfo  {journal}
  {Mon. Not. Roy. Astron. Soc.}\ }\textbf {\bibinfo {volume} {413}},\ \bibinfo
  {pages} {1717} (\bibinfo {year} {2011})},\ \Eprint
  {http://arxiv.org/abs/1010.1519} {arXiv:1010.1519 [astro-ph.CO]} \BibitemShut
  {NoStop}%
\bibitem [{\citenamefont {Fixsen}\ \emph {et~al.}(1996)\citenamefont {Fixsen},
  \citenamefont {Cheng}, \citenamefont {Gales}, \citenamefont {Mather},
  \citenamefont {Shafer},\ and\ \citenamefont {Wright}}]{Fixsen:1996nj}%
  \BibitemOpen
  \bibfield  {author} {\bibinfo {author} {\bibfnamefont {D.~J.}\ \bibnamefont
  {Fixsen}}, \bibinfo {author} {\bibfnamefont {E.~S.}\ \bibnamefont {Cheng}},
  \bibinfo {author} {\bibfnamefont {J.~M.}\ \bibnamefont {Gales}}, \bibinfo
  {author} {\bibfnamefont {J.~C.}\ \bibnamefont {Mather}}, \bibinfo {author}
  {\bibfnamefont {R.~A.}\ \bibnamefont {Shafer}}, \ and\ \bibinfo {author}
  {\bibfnamefont {E.~L.}\ \bibnamefont {Wright}},\ }\href {\doibase
  10.1086/178173} {\bibfield  {journal} {\bibinfo  {journal} {Astrophys. J.}\
  }\textbf {\bibinfo {volume} {473}},\ \bibinfo {pages} {576} (\bibinfo {year}
  {1996})},\ \Eprint {http://arxiv.org/abs/astro-ph/9605054}
  {arXiv:astro-ph/9605054} \BibitemShut {NoStop}%
\bibitem [{\citenamefont {Musco}\ \emph {et~al.}(2021)\citenamefont {Musco},
  \citenamefont {De~Luca}, \citenamefont {Franciolini},\ and\ \citenamefont
  {Riotto}}]{Musco:2020jjb}%
  \BibitemOpen
  \bibfield  {author} {\bibinfo {author} {\bibfnamefont {I.}~\bibnamefont
  {Musco}}, \bibinfo {author} {\bibfnamefont {V.}~\bibnamefont {De~Luca}},
  \bibinfo {author} {\bibfnamefont {G.}~\bibnamefont {Franciolini}}, \ and\
  \bibinfo {author} {\bibfnamefont {A.}~\bibnamefont {Riotto}},\ }\href
  {\doibase 10.1103/PhysRevD.103.063538} {\bibfield  {journal} {\bibinfo
  {journal} {Phys. Rev. D}\ }\textbf {\bibinfo {volume} {103}},\ \bibinfo
  {pages} {063538} (\bibinfo {year} {2021})},\ \Eprint
  {http://arxiv.org/abs/2011.03014} {arXiv:2011.03014 [astro-ph.CO]}
  \BibitemShut {NoStop}%
\bibitem [{\citenamefont {Salopek}\ and\ \citenamefont
  {Bond}(1990)}]{Salopek:1990jq}%
  \BibitemOpen
  \bibfield  {author} {\bibinfo {author} {\bibfnamefont {D.~S.}\ \bibnamefont
  {Salopek}}\ and\ \bibinfo {author} {\bibfnamefont {J.~R.}\ \bibnamefont
  {Bond}},\ }\href {\doibase 10.1103/PhysRevD.42.3936} {\bibfield  {journal}
  {\bibinfo  {journal} {Phys. Rev. D}\ }\textbf {\bibinfo {volume} {42}},\
  \bibinfo {pages} {3936} (\bibinfo {year} {1990})}\BibitemShut {NoStop}%
\bibitem [{\citenamefont {Deruelle}\ and\ \citenamefont
  {Langlois}(1995)}]{Deruelle:1994iz}%
  \BibitemOpen
  \bibfield  {author} {\bibinfo {author} {\bibfnamefont {N.}~\bibnamefont
  {Deruelle}}\ and\ \bibinfo {author} {\bibfnamefont {D.}~\bibnamefont
  {Langlois}},\ }\href {\doibase 10.1103/PhysRevD.52.2007} {\bibfield
  {journal} {\bibinfo  {journal} {Phys. Rev. D}\ }\textbf {\bibinfo {volume}
  {52}},\ \bibinfo {pages} {2007} (\bibinfo {year} {1995})},\ \Eprint
  {http://arxiv.org/abs/gr-qc/9411040} {arXiv:gr-qc/9411040} \BibitemShut
  {NoStop}%
\bibitem [{\citenamefont {Afshordi}\ and\ \citenamefont
  {Brandenberger}(2001)}]{Afshordi:2000nr}%
  \BibitemOpen
  \bibfield  {author} {\bibinfo {author} {\bibfnamefont {N.}~\bibnamefont
  {Afshordi}}\ and\ \bibinfo {author} {\bibfnamefont {R.~H.}\ \bibnamefont
  {Brandenberger}},\ }\href {\doibase 10.1103/PhysRevD.63.123505} {\bibfield
  {journal} {\bibinfo  {journal} {Phys. Rev. D}\ }\textbf {\bibinfo {volume}
  {63}},\ \bibinfo {pages} {123505} (\bibinfo {year} {2001})},\ \Eprint
  {http://arxiv.org/abs/gr-qc/0011075} {arXiv:gr-qc/0011075} \BibitemShut
  {NoStop}%
\bibitem [{\citenamefont {Shibata}\ and\ \citenamefont
  {Sasaki}(1999)}]{Shibata:1999zs}%
  \BibitemOpen
  \bibfield  {author} {\bibinfo {author} {\bibfnamefont {M.}~\bibnamefont
  {Shibata}}\ and\ \bibinfo {author} {\bibfnamefont {M.}~\bibnamefont
  {Sasaki}},\ }\href {\doibase 10.1103/PhysRevD.60.084002} {\bibfield
  {journal} {\bibinfo  {journal} {Phys. Rev. D}\ }\textbf {\bibinfo {volume}
  {60}},\ \bibinfo {pages} {084002} (\bibinfo {year} {1999})},\ \Eprint
  {http://arxiv.org/abs/gr-qc/9905064} {arXiv:gr-qc/9905064} \BibitemShut
  {NoStop}%
\bibitem [{\citenamefont {Kawasaki}\ and\ \citenamefont
  {Nakatsuka}(2019)}]{Kawasaki:2019mbl}%
  \BibitemOpen
  \bibfield  {author} {\bibinfo {author} {\bibfnamefont {M.}~\bibnamefont
  {Kawasaki}}\ and\ \bibinfo {author} {\bibfnamefont {H.}~\bibnamefont
  {Nakatsuka}},\ }\href {\doibase 10.1103/PhysRevD.99.123501} {\bibfield
  {journal} {\bibinfo  {journal} {Phys. Rev. D}\ }\textbf {\bibinfo {volume}
  {99}},\ \bibinfo {pages} {123501} (\bibinfo {year} {2019})},\ \Eprint
  {http://arxiv.org/abs/1903.02994} {arXiv:1903.02994 [astro-ph.CO]}
  \BibitemShut {NoStop}%
\bibitem [{\citenamefont {Young}\ \emph {et~al.}(2019)\citenamefont {Young},
  \citenamefont {Musco},\ and\ \citenamefont {Byrnes}}]{Young:2019yug}%
  \BibitemOpen
  \bibfield  {author} {\bibinfo {author} {\bibfnamefont {S.}~\bibnamefont
  {Young}}, \bibinfo {author} {\bibfnamefont {I.}~\bibnamefont {Musco}}, \ and\
  \bibinfo {author} {\bibfnamefont {C.~T.}\ \bibnamefont {Byrnes}},\ }\href
  {\doibase 10.1088/1475-7516/2019/11/012} {\bibfield  {journal} {\bibinfo
  {journal} {JCAP}\ }\textbf {\bibinfo {volume} {11}},\ \bibinfo {pages} {012}
  (\bibinfo {year} {2019})},\ \Eprint {http://arxiv.org/abs/1904.00984}
  {arXiv:1904.00984 [astro-ph.CO]} \BibitemShut {NoStop}%
\bibitem [{\citenamefont {Kalaja}\ \emph {et~al.}(2019)\citenamefont {Kalaja},
  \citenamefont {Bellomo}, \citenamefont {Bartolo}, \citenamefont {Bertacca},
  \citenamefont {Matarrese}, \citenamefont {Musco}, \citenamefont
  {Raccanelli},\ and\ \citenamefont {Verde}}]{Kalaja:2019uju}%
  \BibitemOpen
  \bibfield  {author} {\bibinfo {author} {\bibfnamefont {A.}~\bibnamefont
  {Kalaja}}, \bibinfo {author} {\bibfnamefont {N.}~\bibnamefont {Bellomo}},
  \bibinfo {author} {\bibfnamefont {N.}~\bibnamefont {Bartolo}}, \bibinfo
  {author} {\bibfnamefont {D.}~\bibnamefont {Bertacca}}, \bibinfo {author}
  {\bibfnamefont {S.}~\bibnamefont {Matarrese}}, \bibinfo {author}
  {\bibfnamefont {I.}~\bibnamefont {Musco}}, \bibinfo {author} {\bibfnamefont
  {A.}~\bibnamefont {Raccanelli}}, \ and\ \bibinfo {author} {\bibfnamefont
  {L.}~\bibnamefont {Verde}},\ }\href {\doibase 10.1088/1475-7516/2019/10/031}
  {\bibfield  {journal} {\bibinfo  {journal} {JCAP}\ }\textbf {\bibinfo
  {volume} {10}},\ \bibinfo {pages} {031} (\bibinfo {year} {2019})},\ \Eprint
  {http://arxiv.org/abs/1908.03596} {arXiv:1908.03596 [astro-ph.CO]}
  \BibitemShut {NoStop}%
\bibitem [{\citenamefont {Yoo}\ \emph {et~al.}(2018)\citenamefont {Yoo},
  \citenamefont {Harada}, \citenamefont {Garriga},\ and\ \citenamefont
  {Kohri}}]{Yoo:2018kvb}%
  \BibitemOpen
  \bibfield  {author} {\bibinfo {author} {\bibfnamefont {C.-M.}\ \bibnamefont
  {Yoo}}, \bibinfo {author} {\bibfnamefont {T.}~\bibnamefont {Harada}},
  \bibinfo {author} {\bibfnamefont {J.}~\bibnamefont {Garriga}}, \ and\
  \bibinfo {author} {\bibfnamefont {K.}~\bibnamefont {Kohri}},\ }\href
  {\doibase 10.1093/ptep/pty120} {\bibfield  {journal} {\bibinfo  {journal}
  {PTEP}\ }\textbf {\bibinfo {volume} {2018}},\ \bibinfo {pages} {123E01}
  (\bibinfo {year} {2018})},\ \Eprint {http://arxiv.org/abs/1805.03946}
  {arXiv:1805.03946 [astro-ph.CO]} \BibitemShut {NoStop}%
\bibitem [{\citenamefont {De~Luca}\ \emph {et~al.}(2019)\citenamefont
  {De~Luca}, \citenamefont {Franciolini}, \citenamefont {Kehagias},
  \citenamefont {Peloso}, \citenamefont {Riotto},\ and\ \citenamefont
  {\"Unal}}]{DeLuca:2019qsy}%
  \BibitemOpen
  \bibfield  {author} {\bibinfo {author} {\bibfnamefont {V.}~\bibnamefont
  {De~Luca}}, \bibinfo {author} {\bibfnamefont {G.}~\bibnamefont
  {Franciolini}}, \bibinfo {author} {\bibfnamefont {A.}~\bibnamefont
  {Kehagias}}, \bibinfo {author} {\bibfnamefont {M.}~\bibnamefont {Peloso}},
  \bibinfo {author} {\bibfnamefont {A.}~\bibnamefont {Riotto}}, \ and\ \bibinfo
  {author} {\bibfnamefont {C.}~\bibnamefont {\"Unal}},\ }\href {\doibase
  10.1088/1475-7516/2019/07/048} {\bibfield  {journal} {\bibinfo  {journal}
  {JCAP}\ }\textbf {\bibinfo {volume} {07}},\ \bibinfo {pages} {048} (\bibinfo
  {year} {2019})},\ \Eprint {http://arxiv.org/abs/1904.00970} {arXiv:1904.00970
  [astro-ph.CO]} \BibitemShut {NoStop}%
\bibitem [{\citenamefont {Gow}\ \emph {et~al.}(2021)\citenamefont {Gow},
  \citenamefont {Byrnes}, \citenamefont {Cole},\ and\ \citenamefont
  {Young}}]{Gow:2020bzo}%
  \BibitemOpen
  \bibfield  {author} {\bibinfo {author} {\bibfnamefont {A.~D.}\ \bibnamefont
  {Gow}}, \bibinfo {author} {\bibfnamefont {C.~T.}\ \bibnamefont {Byrnes}},
  \bibinfo {author} {\bibfnamefont {P.~S.}\ \bibnamefont {Cole}}, \ and\
  \bibinfo {author} {\bibfnamefont {S.}~\bibnamefont {Young}},\ }\href
  {\doibase 10.1088/1475-7516/2021/02/002} {\bibfield  {journal} {\bibinfo
  {journal} {JCAP}\ }\textbf {\bibinfo {volume} {02}},\ \bibinfo {pages} {002}
  (\bibinfo {year} {2021})},\ \Eprint {http://arxiv.org/abs/2008.03289}
  {arXiv:2008.03289 [astro-ph.CO]} \BibitemShut {NoStop}%
\bibitem [{\citenamefont {Germani}\ and\ \citenamefont
  {Musco}(2019)}]{Germani:2018jgr}%
  \BibitemOpen
  \bibfield  {author} {\bibinfo {author} {\bibfnamefont {C.}~\bibnamefont
  {Germani}}\ and\ \bibinfo {author} {\bibfnamefont {I.}~\bibnamefont
  {Musco}},\ }\href {\doibase 10.1103/PhysRevLett.122.141302} {\bibfield
  {journal} {\bibinfo  {journal} {Phys. Rev. Lett.}\ }\textbf {\bibinfo
  {volume} {122}},\ \bibinfo {pages} {141302} (\bibinfo {year} {2019})},\
  \Eprint {http://arxiv.org/abs/1805.04087} {arXiv:1805.04087 [astro-ph.CO]}
  \BibitemShut {NoStop}%
\bibitem [{\citenamefont {Musco}(2019)}]{Musco:2018rwt}%
  \BibitemOpen
  \bibfield  {author} {\bibinfo {author} {\bibfnamefont {I.}~\bibnamefont
  {Musco}},\ }\href {\doibase 10.1103/PhysRevD.100.123524} {\bibfield
  {journal} {\bibinfo  {journal} {Phys. Rev. D}\ }\textbf {\bibinfo {volume}
  {100}},\ \bibinfo {pages} {123524} (\bibinfo {year} {2019})},\ \Eprint
  {http://arxiv.org/abs/1809.02127} {arXiv:1809.02127 [gr-qc]} \BibitemShut
  {NoStop}%
\bibitem [{\citenamefont {Sasaki}\ \emph {et~al.}(2018)\citenamefont {Sasaki},
  \citenamefont {Suyama}, \citenamefont {Tanaka},\ and\ \citenamefont
  {Yokoyama}}]{Sasaki:2018dmp}%
  \BibitemOpen
  \bibfield  {author} {\bibinfo {author} {\bibfnamefont {M.}~\bibnamefont
  {Sasaki}}, \bibinfo {author} {\bibfnamefont {T.}~\bibnamefont {Suyama}},
  \bibinfo {author} {\bibfnamefont {T.}~\bibnamefont {Tanaka}}, \ and\ \bibinfo
  {author} {\bibfnamefont {S.}~\bibnamefont {Yokoyama}},\ }\href {\doibase
  10.1088/1361-6382/aaa7b4} {\bibfield  {journal} {\bibinfo  {journal} {Class.
  Quant. Grav.}\ }\textbf {\bibinfo {volume} {35}},\ \bibinfo {pages} {063001}
  (\bibinfo {year} {2018})},\ \Eprint {http://arxiv.org/abs/1801.05235}
  {arXiv:1801.05235 [astro-ph.CO]} \BibitemShut {NoStop}%
\bibitem [{\citenamefont {Press}\ and\ \citenamefont
  {Schechter}(1974)}]{Press:1973iz}%
  \BibitemOpen
  \bibfield  {author} {\bibinfo {author} {\bibfnamefont {W.~H.}\ \bibnamefont
  {Press}}\ and\ \bibinfo {author} {\bibfnamefont {P.}~\bibnamefont
  {Schechter}},\ }\href {\doibase 10.1086/152650} {\bibfield  {journal}
  {\bibinfo  {journal} {Astrophys. J.}\ }\textbf {\bibinfo {volume} {187}},\
  \bibinfo {pages} {425} (\bibinfo {year} {1974})}\BibitemShut {NoStop}%
\bibitem [{\citenamefont {Ananda}\ \emph {et~al.}(2007)\citenamefont {Ananda},
  \citenamefont {Clarkson},\ and\ \citenamefont {Wands}}]{Ananda:2006af}%
  \BibitemOpen
  \bibfield  {author} {\bibinfo {author} {\bibfnamefont {K.~N.}\ \bibnamefont
  {Ananda}}, \bibinfo {author} {\bibfnamefont {C.}~\bibnamefont {Clarkson}}, \
  and\ \bibinfo {author} {\bibfnamefont {D.}~\bibnamefont {Wands}},\ }\href
  {\doibase 10.1103/PhysRevD.75.123518} {\bibfield  {journal} {\bibinfo
  {journal} {Phys. Rev. D}\ }\textbf {\bibinfo {volume} {75}},\ \bibinfo
  {pages} {123518} (\bibinfo {year} {2007})},\ \Eprint
  {http://arxiv.org/abs/gr-qc/0612013} {arXiv:gr-qc/0612013} \BibitemShut
  {NoStop}%
\bibitem [{\citenamefont {Baumann}\ \emph {et~al.}(2007)\citenamefont
  {Baumann}, \citenamefont {Steinhardt}, \citenamefont {Takahashi},\ and\
  \citenamefont {Ichiki}}]{Baumann:2007zm}%
  \BibitemOpen
  \bibfield  {author} {\bibinfo {author} {\bibfnamefont {D.}~\bibnamefont
  {Baumann}}, \bibinfo {author} {\bibfnamefont {P.~J.}\ \bibnamefont
  {Steinhardt}}, \bibinfo {author} {\bibfnamefont {K.}~\bibnamefont
  {Takahashi}}, \ and\ \bibinfo {author} {\bibfnamefont {K.}~\bibnamefont
  {Ichiki}},\ }\href {\doibase 10.1103/PhysRevD.76.084019} {\bibfield
  {journal} {\bibinfo  {journal} {Phys. Rev. D}\ }\textbf {\bibinfo {volume}
  {76}},\ \bibinfo {pages} {084019} (\bibinfo {year} {2007})},\ \Eprint
  {http://arxiv.org/abs/hep-th/0703290} {arXiv:hep-th/0703290} \BibitemShut
  {NoStop}%
\bibitem [{\citenamefont {Dom\`enech}(2021)}]{Domenech:2021ztg}%
  \BibitemOpen
  \bibfield  {author} {\bibinfo {author} {\bibfnamefont {G.}~\bibnamefont
  {Dom\`enech}},\ }\href {\doibase 10.3390/universe7110398} {\bibfield
  {journal} {\bibinfo  {journal} {Universe}\ }\textbf {\bibinfo {volume} {7}},\
  \bibinfo {pages} {398} (\bibinfo {year} {2021})},\ \Eprint
  {http://arxiv.org/abs/2109.01398} {arXiv:2109.01398 [gr-qc]} \BibitemShut
  {NoStop}%
\bibitem [{\citenamefont {Janssen}\ \emph {et~al.}(2015)\citenamefont {Janssen}
  \emph {et~al.}}]{Janssen:2014dka}%
  \BibitemOpen
  \bibfield  {author} {\bibinfo {author} {\bibfnamefont {G.}~\bibnamefont
  {Janssen}} \emph {et~al.},\ }\href {\doibase 10.22323/1.215.0037} {\bibfield
  {journal} {\bibinfo  {journal} {PoS}\ }\textbf {\bibinfo {volume}
  {AASKA14}},\ \bibinfo {pages} {037} (\bibinfo {year} {2015})},\ \Eprint
  {http://arxiv.org/abs/1501.00127} {arXiv:1501.00127 [astro-ph.IM]}
  \BibitemShut {NoStop}%
\bibitem [{\citenamefont {Ruan}\ \emph {et~al.}(2020)\citenamefont {Ruan},
  \citenamefont {Guo}, \citenamefont {Cai},\ and\ \citenamefont
  {Zhang}}]{Ruan:2018tsw}%
  \BibitemOpen
  \bibfield  {author} {\bibinfo {author} {\bibfnamefont {W.-H.}\ \bibnamefont
  {Ruan}}, \bibinfo {author} {\bibfnamefont {Z.-K.}\ \bibnamefont {Guo}},
  \bibinfo {author} {\bibfnamefont {R.-G.}\ \bibnamefont {Cai}}, \ and\
  \bibinfo {author} {\bibfnamefont {Y.-Z.}\ \bibnamefont {Zhang}},\ }\href
  {\doibase 10.1142/S0217751X2050075X} {\bibfield  {journal} {\bibinfo
  {journal} {Int. J. Mod. Phys. A}\ }\textbf {\bibinfo {volume} {35}},\
  \bibinfo {pages} {2050075} (\bibinfo {year} {2020})},\ \Eprint
  {http://arxiv.org/abs/1807.09495} {arXiv:1807.09495 [gr-qc]} \BibitemShut
  {NoStop}%
\bibitem [{\citenamefont {Kawamura}\ \emph {et~al.}(2011)\citenamefont
  {Kawamura} \emph {et~al.}}]{Kawamura:2011zz}%
  \BibitemOpen
  \bibfield  {author} {\bibinfo {author} {\bibfnamefont {S.}~\bibnamefont
  {Kawamura}} \emph {et~al.},\ }\href {\doibase 10.1088/0264-9381/28/9/094011}
  {\bibfield  {journal} {\bibinfo  {journal} {Class. Quant. Grav.}\ }\textbf
  {\bibinfo {volume} {28}},\ \bibinfo {pages} {094011} (\bibinfo {year}
  {2011})}\BibitemShut {NoStop}%
\bibitem [{\citenamefont {Crowder}\ and\ \citenamefont
  {Cornish}(2005)}]{Crowder:2005nr}%
  \BibitemOpen
  \bibfield  {author} {\bibinfo {author} {\bibfnamefont {J.}~\bibnamefont
  {Crowder}}\ and\ \bibinfo {author} {\bibfnamefont {N.~J.}\ \bibnamefont
  {Cornish}},\ }\href {\doibase 10.1103/PhysRevD.72.083005} {\bibfield
  {journal} {\bibinfo  {journal} {Phys. Rev. D}\ }\textbf {\bibinfo {volume}
  {72}},\ \bibinfo {pages} {083005} (\bibinfo {year} {2005})},\ \Eprint
  {http://arxiv.org/abs/gr-qc/0506015} {arXiv:gr-qc/0506015} \BibitemShut
  {NoStop}%
\bibitem [{\citenamefont {Baker}\ \emph {et~al.}(2019)\citenamefont {Baker}
  \emph {et~al.}}]{Baker:2019nia}%
  \BibitemOpen
  \bibfield  {author} {\bibinfo {author} {\bibfnamefont {J.}~\bibnamefont
  {Baker}} \emph {et~al.},\ }\href@noop {} {\  (\bibinfo {year} {2019})},\
  \Eprint {http://arxiv.org/abs/1907.06482} {arXiv:1907.06482 [astro-ph.IM]}
  \BibitemShut {NoStop}%
\bibitem [{\citenamefont {El-Neaj}\ \emph {et~al.}(2020)\citenamefont {El-Neaj}
  \emph {et~al.}}]{AEDGE:2019nxb}%
  \BibitemOpen
  \bibfield  {author} {\bibinfo {author} {\bibfnamefont {Y.~A.}\ \bibnamefont
  {El-Neaj}} \emph {et~al.} (\bibinfo {collaboration} {AEDGE}),\ }\href
  {\doibase 10.1140/epjqt/s40507-020-0080-0} {\bibfield  {journal} {\bibinfo
  {journal} {EPJ Quant. Technol.}\ }\textbf {\bibinfo {volume} {7}},\ \bibinfo
  {pages} {6} (\bibinfo {year} {2020})},\ \Eprint
  {http://arxiv.org/abs/1908.00802} {arXiv:1908.00802 [gr-qc]} \BibitemShut
  {NoStop}%
\bibitem [{\citenamefont {Garcia-Bellido}\ \emph {et~al.}(2021)\citenamefont
  {Garcia-Bellido}, \citenamefont {Murayama},\ and\ \citenamefont
  {White}}]{Garcia-Bellido:2021zgu}%
  \BibitemOpen
  \bibfield  {author} {\bibinfo {author} {\bibfnamefont {J.}~\bibnamefont
  {Garcia-Bellido}}, \bibinfo {author} {\bibfnamefont {H.}~\bibnamefont
  {Murayama}}, \ and\ \bibinfo {author} {\bibfnamefont {G.}~\bibnamefont
  {White}},\ }\href {\doibase 10.1088/1475-7516/2021/12/023} {\bibfield
  {journal} {\bibinfo  {journal} {JCAP}\ }\textbf {\bibinfo {volume} {12}},\
  \bibinfo {pages} {023} (\bibinfo {year} {2021})},\ \Eprint
  {http://arxiv.org/abs/2104.04778} {arXiv:2104.04778 [hep-ph]} \BibitemShut
  {NoStop}%
\bibitem [{\citenamefont {Sesana}\ \emph {et~al.}(2021)\citenamefont {Sesana}
  \emph {et~al.}}]{Sesana:2019vho}%
  \BibitemOpen
  \bibfield  {author} {\bibinfo {author} {\bibfnamefont {A.}~\bibnamefont
  {Sesana}} \emph {et~al.},\ }\href {\doibase 10.1007/s10686-021-09709-9}
  {\bibfield  {journal} {\bibinfo  {journal} {Exper. Astron.}\ }\textbf
  {\bibinfo {volume} {51}},\ \bibinfo {pages} {1333} (\bibinfo {year}
  {2021})},\ \Eprint {http://arxiv.org/abs/1908.11391} {arXiv:1908.11391
  [astro-ph.IM]} \BibitemShut {NoStop}%
\bibitem [{\citenamefont {Kohri}\ and\ \citenamefont
  {Terada}(2018)}]{Kohri:2018awv}%
  \BibitemOpen
  \bibfield  {author} {\bibinfo {author} {\bibfnamefont {K.}~\bibnamefont
  {Kohri}}\ and\ \bibinfo {author} {\bibfnamefont {T.}~\bibnamefont {Terada}},\
  }\href {\doibase 10.1103/PhysRevD.97.123532} {\bibfield  {journal} {\bibinfo
  {journal} {Phys. Rev. D}\ }\textbf {\bibinfo {volume} {97}},\ \bibinfo
  {pages} {123532} (\bibinfo {year} {2018})},\ \Eprint
  {http://arxiv.org/abs/1804.08577} {arXiv:1804.08577 [gr-qc]} \BibitemShut
  {NoStop}%
\bibitem [{\citenamefont {Espinosa}\ \emph {et~al.}(2018)\citenamefont
  {Espinosa}, \citenamefont {Racco},\ and\ \citenamefont
  {Riotto}}]{Espinosa:2018eve}%
  \BibitemOpen
  \bibfield  {author} {\bibinfo {author} {\bibfnamefont {J.~R.}\ \bibnamefont
  {Espinosa}}, \bibinfo {author} {\bibfnamefont {D.}~\bibnamefont {Racco}}, \
  and\ \bibinfo {author} {\bibfnamefont {A.}~\bibnamefont {Riotto}},\ }\href
  {\doibase 10.1088/1475-7516/2018/09/012} {\bibfield  {journal} {\bibinfo
  {journal} {JCAP}\ }\textbf {\bibinfo {volume} {09}},\ \bibinfo {pages} {012}
  (\bibinfo {year} {2018})},\ \Eprint {http://arxiv.org/abs/1804.07732}
  {arXiv:1804.07732 [hep-ph]} \BibitemShut {NoStop}%
\bibitem [{\citenamefont {Brill}\ and\ \citenamefont
  {Hartle}(1964)}]{Brill:1964zz}%
  \BibitemOpen
  \bibfield  {author} {\bibinfo {author} {\bibfnamefont {D.~R.}\ \bibnamefont
  {Brill}}\ and\ \bibinfo {author} {\bibfnamefont {J.~B.}\ \bibnamefont
  {Hartle}},\ }\href {\doibase 10.1103/PhysRev.135.B271} {\bibfield  {journal}
  {\bibinfo  {journal} {Phys. Rev.}\ }\textbf {\bibinfo {volume} {135}},\
  \bibinfo {pages} {B271} (\bibinfo {year} {1964})}\BibitemShut {NoStop}%
\bibitem [{\citenamefont {Isaacson}(1968{\natexlab{a}})}]{Isaacson:1968hbi}%
  \BibitemOpen
  \bibfield  {author} {\bibinfo {author} {\bibfnamefont {R.~A.}\ \bibnamefont
  {Isaacson}},\ }\href {\doibase 10.1103/PhysRev.166.1263} {\bibfield
  {journal} {\bibinfo  {journal} {Phys. Rev.}\ }\textbf {\bibinfo {volume}
  {166}},\ \bibinfo {pages} {1263} (\bibinfo {year}
  {1968}{\natexlab{a}})}\BibitemShut {NoStop}%
\bibitem [{\citenamefont {Isaacson}(1968{\natexlab{b}})}]{Isaacson:1968zza}%
  \BibitemOpen
  \bibfield  {author} {\bibinfo {author} {\bibfnamefont {R.~A.}\ \bibnamefont
  {Isaacson}},\ }\href {\doibase 10.1103/PhysRev.166.1272} {\bibfield
  {journal} {\bibinfo  {journal} {Phys. Rev.}\ }\textbf {\bibinfo {volume}
  {166}},\ \bibinfo {pages} {1272} (\bibinfo {year}
  {1968}{\natexlab{b}})}\BibitemShut {NoStop}%
\bibitem [{\citenamefont {Ford}\ and\ \citenamefont
  {Parker}(1977)}]{Ford:1977dj}%
  \BibitemOpen
  \bibfield  {author} {\bibinfo {author} {\bibfnamefont {L.~H.}\ \bibnamefont
  {Ford}}\ and\ \bibinfo {author} {\bibfnamefont {L.}~\bibnamefont {Parker}},\
  }\href {\doibase 10.1103/PhysRevD.16.1601} {\bibfield  {journal} {\bibinfo
  {journal} {Phys. Rev. D}\ }\textbf {\bibinfo {volume} {16}},\ \bibinfo
  {pages} {1601} (\bibinfo {year} {1977})}\BibitemShut {NoStop}%
\bibitem [{\citenamefont {Ota}\ \emph {et~al.}(2022)\citenamefont {Ota},
  \citenamefont {Macpherson},\ and\ \citenamefont {Coulton}}]{Ota:2021fdv}%
  \BibitemOpen
  \bibfield  {author} {\bibinfo {author} {\bibfnamefont {A.}~\bibnamefont
  {Ota}}, \bibinfo {author} {\bibfnamefont {H.~J.}\ \bibnamefont {Macpherson}},
  \ and\ \bibinfo {author} {\bibfnamefont {W.~R.}\ \bibnamefont {Coulton}},\
  }\href {\doibase 10.1103/PhysRevD.106.063521} {\bibfield  {journal} {\bibinfo
   {journal} {Phys. Rev. D}\ }\textbf {\bibinfo {volume} {106}},\ \bibinfo
  {pages} {063521} (\bibinfo {year} {2022})},\ \Eprint
  {http://arxiv.org/abs/2111.09163} {arXiv:2111.09163 [gr-qc]} \BibitemShut
  {NoStop}%
\bibitem [{\citenamefont {Pi}\ and\ \citenamefont {Sasaki}(2020)}]{Pi:2020otn}%
  \BibitemOpen
  \bibfield  {author} {\bibinfo {author} {\bibfnamefont {S.}~\bibnamefont
  {Pi}}\ and\ \bibinfo {author} {\bibfnamefont {M.}~\bibnamefont {Sasaki}},\
  }\href {\doibase 10.1088/1475-7516/2020/09/037} {\bibfield  {journal}
  {\bibinfo  {journal} {JCAP}\ }\textbf {\bibinfo {volume} {09}},\ \bibinfo
  {pages} {037} (\bibinfo {year} {2020})},\ \Eprint
  {http://arxiv.org/abs/2005.12306} {arXiv:2005.12306 [gr-qc]} \BibitemShut
  {NoStop}%
\end{thebibliography}%

\end{document}